\documentclass[12pt,draftclsnofoot,onecolumn,comsoc]{IEEEtran}

\usepackage{xspace,amsmath,amssymb,amsfonts,epsfig,syntonly,mathtools}
\usepackage{cite,bm,color,url,textcomp,empheq,boxedminipage}
\usepackage{algorithmicx,algorithm,algpseudocode}
\usepackage{epstopdf}
\usepackage{empheq}
\usepackage{graphicx,graphics,subfigure}  
\usepackage{multirow,multicol}
\usepackage{psfrag}    
\usepackage{stfloats}

\newtheorem{definition}{\underline{Definition}}

\newtheorem{lemma}{\underline{Lemma}}

\newtheorem{proposition}{\underline{Proposition}}
\newtheorem{example}{\underline{Example}}
\newtheorem{remark}{\underline{Remark}}

\DeclareMathOperator*{\argmin}{arg\,min}

\long\def\symbolfootnote[#1]#2{\begingroup
\def\thefootnote{\fnsymbol{footnote}}
\footnote[#1]{#2}\endgroup}

\psfull

\usepackage[T1]{fontenc}
\usepackage{ifpdf}
 \ifpdf
 \else
 \fi
\begin{document}
\title{Multi-Level Over-the-Air Aggregation of Mobile Edge Computing over D2D \\ Wireless Networks}

\author{Feng~Wang,~\IEEEmembership{Member,~IEEE,}~and~Vincent~K.~N.~Lau,~\IEEEmembership{Fellow,~IEEE}
\thanks{F. Wang is with the School of Information Engineering, Guangdong University of Technology, Guangzhou 510006, China, and also with the Department of Electronic and Computer Engineering, The Hong Kong University of Science and Technology, Hong Kong (e-mail: fengwang13@gdut.edu.cn).}
\thanks{V. K. N. Lau is with the Department of Electronic and Computer Engineering, The Hong Kong University of Science and Technology, Hong Kong (e-mail: eeknlau@ust.hk).}
\vspace{-1.8cm}
}

\maketitle

\begin{abstract}
\vspace{-0.4cm}
In this paper, we consider a wireless multihop device-to-device (D2D) based mobile edge computing (MEC) system, where the destination wireless device (WD) is scheduled to compute nomographic functions. Under the MapReduce framework and motivated by reducing communication resource overhead, we propose a new multi-level over-the-air (OTA) aggregation scheme for the destination WD to collect the individual partially aggregated intermediate values (IVAs) for reduction from multiple source WDs in the data shuffling phase. For OTA aggregation per level, the source WDs employ a channel-inverse structure multiplied by their individual transmit coefficients in transmission over the same time-frequency resource blocks, and the destination WD finally uses a receive filtering factor to construct the aggregated IVA. Under this setup, we develop a unified transceiver design framework that minimizes the mean squared error (MSE) of the aggregated IVA at the destination WD subject to the source WDs' individual power constraints, by jointly optimizing the source WDs' individual transmit coefficients and the destination WD's receive filtering factor. The formulated power-constrained MSE minimization problem is non-convex. First, based on the primal decomposition method, we derive the closed-form solution under the special case of a common transmit coefficient. It shows that all the source WDs' common transmit is determined by the minimal transmit power budget among the source WDs. Next, for the general case, we transform the original problem into a quadratic fractional programming problem, and then develop a low-complexity algorithm to obtain the (near-) optimal solution by leveraging Dinkelbach's algorithm along with the Gaussian randomization method. Numerical results are provided to demonstrate the significant performance gains achieved by the proposed multi-level OTA aggregation scheme over various existing schemes.
\end{abstract}

\vspace{-0.8cm}
\begin{IEEEkeywords}
\vspace{-0.4cm}
 Mobile edge computing, wireless MapReduce, multi-level over-the-air (OTA) aggregation, multihop D2D communications, transceiver optimization.
\end{IEEEkeywords}

\IEEEpeerreviewmaketitle

\section{Introduction}
 With the advancements in latency-sensitive Internet of things (IoT) and artificial intelligence (AI) applications (such as augmented/virtual reality, object recognition, and navigation), mobile edge computing (MEC) has emerged as a promising technique to enable wireless devices (WDs) to have proximity access to cloud-like computing functionalities at the network edge\cite{Feng18-TWC,JunZhang17}. Unlike the centralized cloud computing that stores and processes WDs' data in remote datacenters, MEC pools and exploits the abundant resources distributed at massive edge nodes (such as access points, base stations, routers, smartphones, and tablets), thereby achieving an enhanced capability to provide much more agile, resilient, resource-efficient, scalable, and low-latency services for end-users.

 Currently, a distributed computing framework known as MapReduce\cite{Dean08-ACM} has become prevalent for MEC to enable parallelization and distribution of large-scale computations. The MapReduce framework typically consists of Map, Shuffle, and Reduce phases. Specifically, a collection of WDs individually {\em map} a set of input data files and calculate intermediate values (IVAs) in the Map phase. These IVAs are then exchanged among the WDs in the Shuffle phase such that each WD can obtain the necessary IVAs to finally {\em reduce} a set of output functions in the Reduce phase. In wireless MapReduce systems, multiple WDs can communicate with each other to exchange their IVAs via device-to-device (D2D) communication links or via an access point (or relay). Dedicated radio resources (e.g., frequency channel or time slots) per each IVA flow will be required to mitigate the interference among the WDs. This is highly undesirable as the number of radio resource blocks required will scale in ${\cal O}(NK)$ with the number of WDs $K$ and the number of input data files $N$. To enhance the spectral efficiency in the IVA exchange, some prior works have considered a fixed pool of radio resources shared by all the WDs and optimized the scheduling of the IVA exchange under the radio resource constraints\cite{Li18-TIT,Dean08-ACM}. However, the latency will be increased substantially as the number of WDs increases. In \cite{Li17-TN} and \cite{Li19-TIT}, coded distributed computing (CDC) schemes were investigated for wireless MapReduce systems. In these schemes, coded multicasting methods were employed to reduce the communication load in the Shuffle phase, but at the expense of assigning redundant computation tasks at each WD in the Map phase.

 Recently, over-the-air (OTA) aggregation (a.k.a, over-the-air computation (AirComp)) methods \cite{Nazer07-TIT,Golden13-TCOM,Zhu19-TIT} have been proposed to directly compute a {\em nomographic} function with only ${\cal O}(1)$ radio resource blocks, which is independent of the number of WDs $K$ and the number of input data files $N$. By exploiting the signal superposition property of the wireless multiple access channel (MAC), different WDs can simultaneously transmit their data to the destination node over the same frequency band, and the destination node can then reconstruct the targeted nomographic function value from the received signal directly\cite{Cai18-TSP}. Different from the traditional separated communication-computation design approach, the OTA aggregation method integrates communication and computation as well as achieving a higher spectral efficiency. The OTA aggregation method has attracted extensive research interest in various applications, such as distributed estimation in wireless sensor networks\cite{Golden13-TSP,Zhu19-IOTJ}, gradient aggregation in federated learning\cite{Yang20-TWC,Guo21-IOTJ}, and IoT networks\cite{Chen18-IOTJ,Jie20-TWC,Liu20-TWC}.

 In this paper, we focus on exploiting OTA aggregation to enhance the spectral efficiency in the exchange of IVAs to compute reduce functions over a wireless network with D2D communications. We consider a wireless MapReduce system with multiple WDs, as shown in Fig.~\ref{fig:sys_model}. These WDs can communicate with each other via direct or multihop D2D communication links. The WDs {\em collaboratively} compute a nomographic function as a target output function of the destination WD (i.e., WD $K$). In this case, the reduce function becomes a function of linearly aggregated IVAs. In the Map phase, each WD first calculates the individual IVAs based on the local data files, and then generates a partially-aggregated IVA. By employing multi-level OTA aggregation, the source WDs are enabled to send their partially-aggregated IVAs to the destination WD over the air, and this destination WD estimates the desirable aggregated IVA from the receive signals so as to perform the computation of the reduce function in the Reduce phase. Note that due to the dual effect of multihop D2D communication channel fading and receive noise, it is highly important to implement a unified transceiver design by jointly optimizing the pre-processing operation (e.g., transmit phase/power control) at transmitting WDs and the post-processing operation (e.g., receive filter) at the receiving WD. The following summarizes the main results of the paper.

\begin{itemize}
\item {\bf Multi-Level OTA Aggregation for Collecting Aggregated IVAs:} In order to improve the spectral efficiency of IVA exchanging in MapReduce when computing nomographic functions over multihop D2D communication systems, we propose a novel multi-level OTA aggregation design scheme to collect aggregated IVAs from multiple source WDs for the destination WD. The multihop D2D wireless network under consideration is modeled as a minimum spanning tree (MST) of a connected graph. For OTA aggregation per level, the source WDs employ a channel-inverse structure multiplied by their individual transmit coefficients in transmission over the same time-frequency resource blocks, and the destination WD finally adopts a receiver filter to reconstruct the aggregated IVA. 

\item{\bf Optimized Joint Transceiver Design with MSE Minimization:} To measure the distortion of the destination WD's reduce function output, we adopt the mean squared error (MSE) of the input aggregated IVA constructed by the destination WD as a figure of merit. We develop a joint transceiver design framework that minimizes the MSE of interest subject to the source WDs' individual transmission power constraints, by jointly optimizing the source WDs' transmit coefficients and the destination WD's receive filtering factor. Due to the coupling of transceiver design variables, the formulated power-constrained MSE minimization problem is non-convex and is challenging to solve.

\item{\bf Low-Complexity Algorithms and Performance Analysis:} For the special case of a common transmit coefficient for the source WDs, we first derive the closed-form solution for the multi-level OTA aggregation design using the primal decomposition method. It shows the optimal transmit coefficient is biased, but asymptotically unbiased. For the general case with independent transmit coefficients, we transform the original problem into a quadratic fractional programming problem, and obtain the (near-) optimal solution based on the Dinkelbach's algorithm, along with the Gaussian randomization method. Finally, extensive numerical results are provided to gauge the performance of the proposed multi-level OTA aggregation designs with joint transceiver optimization over the state-of-the-art solutions.
\end{itemize}


The remainder of this paper is organized as follows. Section II describes the system model for computing nomographic functions under the distributed MapReduce framework. Section III presents the proposed multi-level OTA aggregation scheme for MapReduce over a multihop D2D wireless network. Section IV formulates the power-constrained MSE problem under consideration. Section V proposes the closed-form solution for the special case with a common transmit coefficient at the source WDs, and Section VI presents the general solution based on Dinkelbach's algorithm. Section VII provides numerical results to evaluate the proposed scheme. The work is concluded in Section VIII.

{\em Notations:} Throughout this paper, boldface lowercase (uppercase) letters indicate vectors (matrices). We use ${\mathbb R}^{N\times M}$ and ${\mathbb C}^{N\times M}$ to denote the set of real-valued and complex-valued $N\times M$ matrices, respectively. The cardinal number of set ${\cal S}$ is denoted as $|{\cal S}|$; ${\cal S}_1\setminus{\cal S}_2$ denotes set subtraction of ${\cal S}_2$ from ${\cal S}_1$. The superscripts $(\cdot)^T$, $(\cdot)^*$, $(\cdot)^H$ denote the transpose, the complex conjugate, and the complex conjugate transpose operations for a vector or scalar, respectively. For a scalar $x$, $|x|$ denotes its absolute value, and ${\rm Re}(x)$ and ${\rm Im}(x)$ denote its real and imaginary components, respectively. The distribution of a circular symmetric complex Gaussian (CSCG) variable $x$ with mean zero and variance $\sigma^2$ is denoted as $x\sim{\cal CN}(0,\sigma^2)$. Finally, ${\mathbb E}[x]$ denotes the expectation operation.


\section{System Model}

\begin{figure}
\centering
  \includegraphics[width=3.6in]{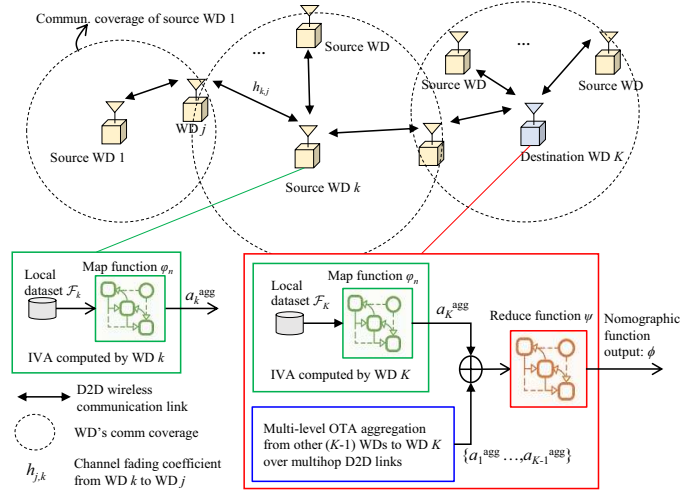}
  \vspace{-0.5cm}
  \caption{System model of MapReduce computing a nomographic output function over a D2D wireless multihop network via multi-level OTA aggregation.}\label{fig:sys_model}
   \vspace{-0.5cm}
\end{figure}

As shown in Fig.~\ref{fig:sys_model}, we consider a wireless MapReduce network consisting of $K$ randomly located WDs. Let ${\cal K}\triangleq\{1,...,K\}$ denote the set of $K$ WDs, which consists of source WDs $\{1,...,K-1\}$ and destination WD $K$. The destination WD $K$ is scheduled to compute a {\em nomographic} function $\phi: \mathbb{C}^{NF\times 1}\mapsto\mathbb{C}$ over a large-scale dataset ${\cal F}\triangleq\{\bm f_1,...,\bm f_N\}$, where $\bm f_n\in\mathbb{C}^{F\times 1}$, $\forall n$. Due to a limited storage capacity, each WD $k\in{\cal K}$ only stores a non-overlapping subset ${\cal F}_k\subseteq{\cal F}$ of data files, such that $\bigcup_{k=1}^K{\cal F}_k={\cal F}$ and ${\cal F}_k\bigcap {\cal F}_j=\emptyset$, $\forall j\neq k$.

For the nomographic function $\phi$, we can always perform the following MapReduce decomposition \cite{Nomographic15-TWC,Buck79-Math}:
\begin{align}\label{eq.nomo}
\underbrace{\phi(\bm f_1,...,\bm f_N)}_{\text{nomographic~func.}}  = \underbrace{\psi}_{\text{reduce~func.}}\Big(\sum_{n=1}^N \underbrace{\varphi_{n}(\bm f_n)}_{\text{map~func.}}\Big) = \psi\Big(\sum_{n=1}^N \underbrace{a_{n}}_{\text{IVA}:~ a_{n}\triangleq \varphi_{n}(\bm f_n)}\Big),
\end{align}
where $\varphi_{n}:\mathbb{C}^{F\times 1}\mapsto\mathbb{C}$ denotes the map function depending only on input data file $\bm f_n$, $a_{n}\triangleq \varphi_{n}(\bm f_n)\in\mathbb{C}$ is defined as the {\em deterministic} IVA generated by the map function $\varphi_{n}$, and $\psi:\mathbb{C}\mapsto\mathbb{C}$ denotes the reduce function that combines the IVAs $\{a_{n}\}_{n=1}^N$ so as to complete the computation of the nomographic function value $\phi$.

\begin{example} [Weighted Average]
The output of an environmental measurement system can be denoted as $\bm d^T \bm f_n>0$ for the measurement data sample vector $\bm f_n$, where $\bm d$ denotes the de-noising vector. As typical nomographic functions, the weight arithmetic average and geometric average of the measurement data over the whole dataset $\{\bm f_1,...,\bm f_N\}$ are respectively expressed as
\begin{align}
\bar{f}_{\text{Arith}}(\bm f_1,...,\bm f_N) = \sum_{n=1}^N \omega_n \bm d^T \bm f_n~~\text{and}~~ \bar{f}_{\text{Geo}}(\bm f_1,...,\bm f_N) = \prod_{n=1}^N (\bm d^T \bm f_n)^{\omega_n},
\end{align}
where $0<\omega_n<1$ denotes the nonnegative weight for $\bm d^T \bm f_n$, $\forall n$, and $\sum_{n=1}^N \omega_n=1$. For the weighted arithmetic average $\bar{f}_{\text{Arith}}$, we have $\psi(x)=1$ and $a_n=\phi_n(f_n)=\omega_n \bm d^T \bm f_n$, $\forall n$. For the weighted geometric average $\bar{f}_{\text{Geo}}$, we have $\psi(x)=\exp(x)$ and $a_n=\phi_n(f_n)=\omega_n \log(\bm d^T \bm f_n)$, $\forall n$.
\end{example}

From the nomographic function decomposition of \eqref{eq.nomo}, it follows that the input variable for the reduce function $\psi$ becomes a sum of local IVAs (i.e., $\sum_{n=1}^N a_{n}$) generated by multiple map functions associated with the nomographic function $\phi$.

\begin{definition}[Partially Aggregated IVA per WD]\label{def:p-a-IVA}
 Based on WD $k$'s local dataset ${\cal F}_k$, we define
 \begin{align}\label{eq.def-agg}
   a_{k}^{\rm agg} \triangleq \sum_{n:\bm f_n\in{\cal F}_k} a_{n},~~\forall k\in{\cal K},
 \end{align}
 as the deterministic partially aggregated IVA of WD $k$ which sums all of its local map function outputs $\{a_n: \bm f_n\in{\cal F}_k\}$, where $k\in{\cal K}$.
\end{definition}

\vspace{-0.4cm}
\subsection{Nomographic Function Computation based on MapReduce Framework}

Based on the nomographic function decomposition in \eqref{eq.nomo} and the MapReduce framework, three consequent phases are required in order for the destination WD $K$ to complete the nomographic function $\phi$, as described in the following.
\begin{itemize}
\item First, in the Map phase, each WD $k\in{\cal K}$ first performs the computation of a number of $|{\cal F}_k|$ map functions $\{\varphi(\bm f_n)\}$, where $f_n\in{\cal F}_k$, and then generates a number of $|{\cal F}_K|$ individual IVAs, i.e., $\{a_k^{\rm agg}\}$. By linearly combining these individual IVAs per WD, a partially aggregated IVA $a_k^{\rm agg}$ is obtained by each WD $k\in{\cal K}$ according to \eqref{eq.def-agg}. Therefore, a total of $K$ partially aggregated IVAs are generated as $\{a^{\rm agg}_k\}_{k=1}^K$ at the $K$ WDs.

\item Second, in the Shuffle phase, each source WD $i\in{\cal K}\setminus\{K\}$ transmits its partially aggregated IVA $a^{\rm agg}_i$ to the destination WD $K$ via direct or multihop wireless D2D communication links (with one or more intermediate WDs serving as relays), such that the destination WD $K$ can collect all the necessary partially aggregated IVAs from the source WDs $\{1,...,K-1\}$ to complete the computation of the reduce function $\psi$. 

\item Third, in the Reduce phase, with the collected partially aggregated IVAs $\{a_1^{\rm agg},...,a_{K-1}^{\rm agg}\}$ from the $(K-1)$ source WDs and its local one $a_K^{\rm agg}$ at hand, the destination WD $K$ proceeds to generate an aggregated IVA $\sum_{k=1}^K a_1^{\rm agg}$ as the input to the reduce function $\psi$. As such, WD $K$ can calculate the targeted nomographic function value $\phi=\psi(\sum_{k=1}^K a_k^{\rm agg})$.
\end{itemize}

\subsection{Graph Representation of Multihop D2D Communication Networks}
In this paper, we assume that all $K$ WDs are synchronized in time so that the time slots are aligned. A common frequency band is shared among all $K$ WDs in both signal transmission and reception. To avoid self-interference, each WD is assumed to operate in a half-duplex communication mode, such that it can only transmit or receive signal at any given time slot\cite{Goldsmith05-book}. As shown in Fig.~\ref{fig:sys_model}, any two distinct WDs can communicate directly with each other via a D2D communication link only when they are within a certain communication range;\footnote{In practice, the communication range/coverage is determined by the minimum signal-to-noise ratio (SNR) level that the destination WD receiver can tolerate.} otherwise, they have to communicate through one or more intermediate WDs serving as relays. Note that such multihop wireless networks are useful in various IoT scenarios when a centralized infrastructure is either unavailable or costly, or when ubiquitous communication services are required without the presence or use of a fixed infrastructure.

To characterize such D2D multihop connectivity among the $K$ WDs, we represent this wireless network as a {\em connected} and {\em undirected} graph ${\cal G}({\cal K},{\cal H})$, where ${\cal K}$ denotes the set of $K$ vertices (or nodes) indexed by the $K$ WDs and ${\cal H}\triangleq \{h_{i,k}\}$ collects all the edges in ${\cal G}({\cal K},{\cal H})$. Each edge $h_{i,k}\in{\cal H}$ represents the D2D complex-valued communication channel fading coefficient from WD $k$ to WD $i$. For the undirected graph ${\cal G}({\cal K},{\cal H})$, it holds that $h_{i,k}\in{\cal H}$ if and only if $h_{k,i}\in{\cal H}$. If $h_{i,k}\in{\cal H}$, then WDs $i$ and $k$ are called {\em adjacent} or {\em neighboring}. For each WD $k\in{\cal K}$, let ${\cal N}_k \triangleq \{i\in{\cal K} | h_{i,k}\in{\cal H}\}$ denotes its neighboring WD set.


\vspace{-0.4cm}
\subsection{Construction of Tree ${\cal T}(K)$ from Graph ${\cal G}({\cal K},{\cal H})$}
In this subsection, we pursue the construction of a minimum spanning tree (MST) ${\cal T}(K)$ from the graph ${\cal G}({\cal K},{\cal H})$, where the shortest pathloss is achieved between the neighboring WDs. 

\begin{definition}[Spanning Tree of ${\cal G}$]
A subgraph of graph ${\cal G}$ is defined as a spanning tree of ${\cal G}$ if it is a tree and contains every vertex of ${\cal G}$. 
\end{definition}

 Denote the inverse of the D2D channel power gain $1/|h_{i,k}|^{2}$ as the weight of each edge $h_{i,k}$ in graph ${\cal G}({\cal K},{\cal H})$. In general, the graph ${\cal G}({\cal K},{\cal H})$ may include many different spanning trees, with destination WD $K$ serving as the root node. The MST of a graph ${\cal G}({\cal K},{\cal H})$ has the minimum weight among all the spanning trees, where the weight of a spanning tree is referred to as the sum of weights of all its edges. Note that the MST of a graph may not be unique, but the MST is provably unique if the edges of this graph have distinct weights. In particular, there are two major greedy algorithms to find an MST ${\cal T}(K)$ from ${\cal G}({\cal K},{\cal H})$, namely, Prim's algorithm and Kruskal's algorithm\cite{Alg2009}. We briefly introduce the two algorithms as follows.
 \begin{itemize}
     \item The basic idea of Prim's algorithm is to grow an MST by adding a new edge in each iteration. We maintain a node set $\cal S$ and an edge set $\cal A$. The MST starts with WD $K$ as the root of the tree by setting ${\cal S}=\{K\}$ and ${\cal A}=\emptyset$. At each iteration, by using, e.g., a priority queue of which each item is a pair $(i,\text{key}[i])$ with $i\in{\cal K}\setminus{\cal S}$ and $\text{key}[i]\triangleq \min_{j\in{\cal S}}{|h_{j,i}|^{-2}}$, one can find the {\em lightest} edge with a minimum weight, such that one edge point is in ${\cal S}$ and the other is in ${\cal K}\setminus{\cal S}$, and then update sets ${\cal A}$ and ${\cal S}$ by adding this edge to ${\cal A}$ and its other endpoint to ${\cal S}$, respectively. If ${\cal V}\setminus{\cal S}=\emptyset$, then the algorithm stops and outputs an MST ${\cal T}(\cal K)$ feathered with $({\cal S},{\cal A})$. For implementation of this Prim's algorithm based on priority queue structures, the total cost is in ${\cal O}((|{\cal H}|+K)\log K)$ time.
  
    \item Unlike Prim's algorithm, Kruskal's algorithm grows a collection of trees and continues until the trees merge into a single MST. In Kruskal's algorithm, the graph is initialized as a collection of trees, each with a single node, and the edges are sorted in a queue based on their weights in a non-decreasing order. In each iteration, the edge with the smallest weight in the queue is selected, and it is checked to see whether it forms a cycle with the spanning tree merged so far. If a cycle is not formed, then we include this edge in the formed tree and remove it from the queue; otherwise, we discard it. If all the nodes are in the same tree or the edge-queue is empty, then the algorithm stops and outputs the MST ${\cal T}(K)$. For implementation of Kruskal's algorithm, the overall time complexity is ${\cal O}(|\cal H|\log K)$.
  \end{itemize}
 
 \begin{figure}
\centering
  \includegraphics[width=3.0in]{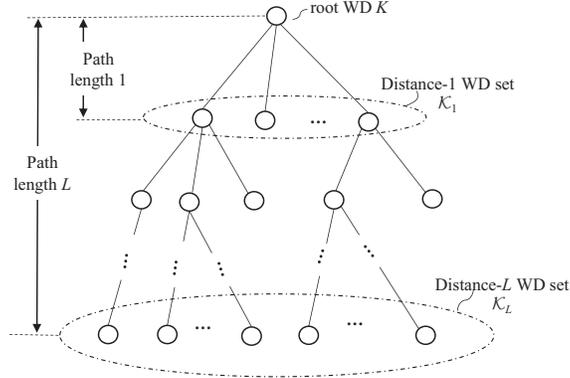}
  \vspace{-0.5cm}
  \caption{An illustration of the MST ${\cal T}(k)$ of graph ${\cal G}({\cal K},{\cal H})$, where the largest path length is $L=\max_{j\in{\cal K}\setminus\{K\}}|{\cal P}_{j\rightarrow K}|$.}\label{fig:Tree_Illustration}
  \vspace{-0.5cm}
\end{figure}

\subsection{Path from Each Source WD $i\in{\cal K}\setminus\{K\}$ to Destination WD $K$}
 Having obtained the MST ${\cal T}(K)$ from  ${\cal G}({\cal K},{\cal H})$, as in Fig.~\ref{fig:Tree_Illustration}, based on either Prim's algorithm or Kruskal's algorithm, we next discuss the (multihop) path from each source WD $i\in{\cal K}\setminus\{K\}$ to the destination WD $K$. For MST ${\cal T}(K)$, an important property is the path uniqueness\cite{Graph-Book}. 

\begin{lemma}[Path Uniqueness\cite{Graph-Book}]\label{lem.path}
For MST ${\cal T}(K)$, any two distinct vertices are connected by exactly one path.
\end{lemma}

Based on Lemma~\ref{lem.path}, the path between any source WD $i\in{\cal K}\setminus\{K\}$ and destination WD $K$ is always unique. Denote by ${\cal P}_{i\rightarrow k}$ the path between WDs $i$ and $k$, where $i\neq k$. Let set ${\cal H}({\cal P}_{i\rightarrow k})$ collect the edges in path ${\cal P}_{i\rightarrow K}$, and the length of ${\cal P}_{i\rightarrow k}$ is defined as its edge number $|{\cal H}({\cal P}_{i\rightarrow k})|$.

\begin{definition}[Distance-$\ell$ Set of WDs] \label{def:distance-set}
For the MST ${\cal T}(K)$, we define set ${\cal K}_{\ell}$ as the distance-$\ell$ set of WDs that are the same length of $\ell$ away from WD $K$, where $\ell=1,...,L$ with $L\triangleq \max_{i\in{\cal K}\setminus\{K\}}|{\cal H}({\cal P}_{i\rightarrow K})|$. In other words, for WD $i\in{\cal K}_{\ell}$, the length of the path between source WD $i$ and destination WD $K$ is exactly equal to $\ell$; i.e., ${\cal K}_{\ell}\triangleq \{i\in{\cal K}\big|~|{\cal H}({\cal P}_{i\rightarrow K})|=\ell \}$.
\end{definition}

\begin{definition}[Direct Parent WD]\label{def:direct-parent}
For a WD $i\in{\cal K}_{\ell}$ in the distance-$\ell$ WD set of tree ${\cal T}(K)$, we define WD $\pi(i)$ as WD $i$'s direct parent WD if this WD $\pi(i)$ belongs to the intersection of WD $i$'s neighboring set ${\cal N}_i$ and distance-$(\ell-1)$ set ${\cal K}_{\ell-1}$, i.e., $\pi(i)\in {\cal N}_i \bigcap {\cal K}_{\ell-1}$, where $\ell=1,...,L$, and we define ${\cal K}_{0}\triangleq \{K\}$ for notational convenience.
\end{definition}

The path uniqueness property in Lemma~\ref{lem.path} yields that each source WD $i\in{\cal K}\setminus\{K\}$ has only one direct parent WD $\pi(i)$. Correspondingly, the intersection set ${\cal N}_i \bigcap {\cal K}_{\ell-1}$ contains only one WD $\pi(i)$ as its element, i.e., ${\cal N}_i \bigcap {\cal K}_{\ell-1}=\{\pi(i)\}$.

\vspace{-0.4cm}
\section{Multi-Level OTA Aggregation for MapReduce over D2D Networks}
Under the constructed MST ${\cal T}(K)$ for MapReduce computing over a D2D network, in this section, we first review the conventional digital method for partially aggregated IVA collection in MapReduce, and then present the proposed analog multi-level OTA aggregation scheme.


\vspace{-0.4cm}
 \subsection{Overview of MapReduce using Conventional Digital Communications}

 In this subsection, we provide an overview of the conventional digital communication scheme to implement the nomographic function computation based on the MapReduce framework in~\eqref{eq.nomo}.

 Let $[0,A]$ denote the range of both the real and imaginary components of the source WDs' partially aggregated IVAs, where $A>0$ is a known parameter decided by the WDs' map function output range. Suppose each source WD $i\in{\cal K}\setminus\{K\}$ quantizes its partially aggregated IVA $a_i^{\rm agg}$ into $Q$ bits to generate a discrete message. This can be achieved by uniformly dividing $[0,A]$ into intervals of length $\Delta=A/(2^Q-1)$, and rounding both ${\rm Re}(a_i^{\rm agg})$ and ${\rm Im}(a_i^{\rm agg})$ to the nearest endpoints of these small intervals\cite{Xiao06-TSP}. Denote by $\{ b^{R}_{i,1},...,b^{R}_{i,Q}\}$ and $\{b^{I}_{i,1},...,b^{I}_{i,Q}\}$ the quantized bits for the real and imaginary components of the partially aggregated IVA $a_i^{\rm agg}$, respectively. The discrete message $m_i$ generated by the quantizer of source WD $i\in{\cal K}\setminus\{K\}$ is\cite{Xiao06-TSP}
 \begin{align}
 m_i = \sum_{i=1}^{Q}  b^{R}_{i,q}2^{-q}\Delta + \sqrt{-1}\times\Big( \sum_{i=1}^{Q}  b^{I}_{i,q}2^{-q}\Delta \Big).
 \end{align}
 
 In order to transmit $m_i$ to destination WD $K$, source WD $i$ needs to send the quantized bits $\{b^{R}_{i,1},...,b^{R}_{i,Q}\}$ and $\{b^{I}_{i,1},...,b^{I}_{i,Q}\}$ via a one- or multi-hop D2D communication link from WD $i$ to WD $K$, e.g., by adopting an uncoded quadrature amplitude modulation (QAM) transmission scheme \cite{Cui05-TWC,Xiao06-TSP}. Denote by $\hat{m}_i$ the estimated version of message $m_i$ at destination WD $K$. Based on $\{\hat{m}_i\}$, WD $K$ proceeds to construct an estimated version of $\sum_{i=1}^{K-1}a_i^{\rm agg}$ as the input of the reduction function $\psi$.

 Consider an uncoded QAM scheme\cite{Cui05-TWC}. Let $s_{R,i}$ and $s_{I,i}$ denote the $2^Q$-QAM symbols mapped by $\{b^{R}_{i,1},...,b^{R}_{i,Q}\}$ and $\{b^{I}_{i,1},...,b^{I}_{i,Q}\}$, respectively. To guarantee a reliable communication, a pool of dedicated radio resource blocks (RBs) is allocated for source WD $i\in{\cal K}\setminus\{K\}$ to transmit symbols $s_{R,i}$ and $s_{I,i}$ to its direct parent WD $\pi(i)$. Denote by $y_{R,\pi(i)}$ and $y_{I,\pi(i)}$ WD $\pi(i)$'s received signals when WD $i$ transmits symbols $s_{R,i}$ and $s_{I,i}$, respectively. Hence, the signal model is
 \begin{align}
 y_{q,\pi(i)} = h_{\pi(i),i} \sqrt{P_i} s_{q,i} + n_{\pi(i)},~~ q\in\{R,I\},
 \end{align}
 where $h_{\pi(i),i}\in\mathbb{C}$ is the channel fading coefficient from WD $i$ to WD $\pi(i)$, $P_i>0$ denotes the source WD $i$'s transmit power, and $n_{\pi(i)}\sim {\cal CN}(0,\sigma^2)$ denotes the additive white Gaussian noise (AWGN) at WD $\pi(i)$'s receiver. Then, for $q\in\{R,I\}$, WD $\pi(i)$ decodes symbol $s_{q,i}$ from $ y_{q,\pi(i)}$, and forwards it to the direct parent WD $\pi_2(i)\triangleq \pi(\pi(i))$ by transmission over a dedicated RB. As such, $2|{\cal H}({\cal P}_{i\rightarrow K})|$ RBs are needed to transmit $s_{R,i}$ and $s_{I,i}$ from WD $i$ to WD $K$ over this D2D multihop network. Therefore, a total of $\sum_{i=1}^{K-1}2|{\cal H}({\cal P}_{i\rightarrow K})|$ RBs are required in the Shuffle phase. In addition, a dedicated pilot per WD $i\in{\cal K}\setminus\{K\}$ is required to estimate a number of channel coefficients $\{h_{\pi(i),i},...,h_{K,\pi_{\ell_i}(i)}\}$ for $\ell_i$-hop transmission from source WD $i$ to destination WD $K$. Hence, the total number of dedicated pilots is ${\cal O}(2KL)$. This means the brute-force digital transmission scheme imposes a high requirement on the radio resources as well as pilot overheads.

 To alleviate the radio resource requirement, all the source WDs in set ${\cal K}\setminus\{K\}$ can share a common pool of radio resources (e.g., a small number of frequency bands). In this case, a multiple access and scheduling protocol is needed to coordinate the channel access of $(K-1)$ source WDs to avoid interference. However, this incurs a significant latency ${\cal O}(K)$.

\vspace{-0.5cm}
\subsection{Multi-Level OTA Aggregation Protocol}

\begin{figure}
\centering
  \includegraphics[width=3.2in]{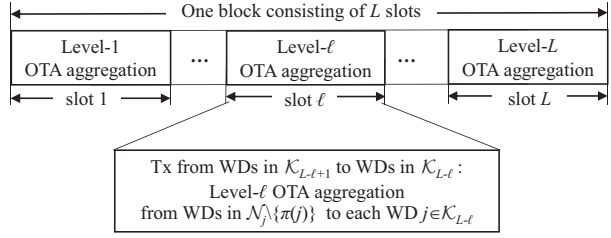}
  \vspace{-0.5cm}
  \caption{Proposed transmission protocol for multi-level OTA aggregation over the multihop wireless D2D network ${\cal G}({\cal K},{\cal H})$, where the block consists of $L$ equal-duration slots.}\label{fig:Tx_Protocol}
  \vspace{-0.5cm}
\end{figure}

\begin{figure}
\centering
  \includegraphics[width=3.2in]{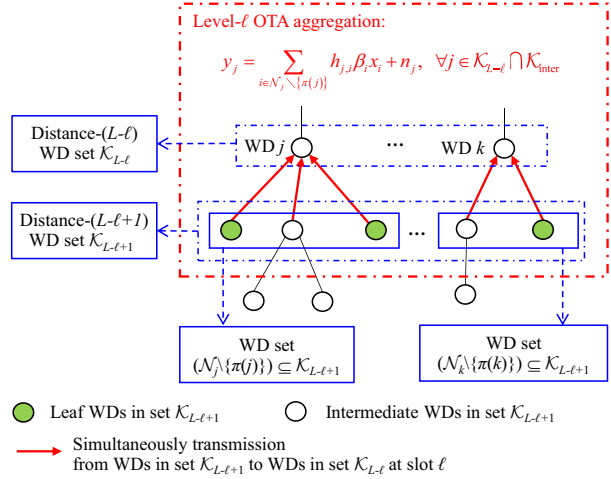}
    \vspace{-0.5cm}
  \caption{An illustration of level-$\ell$ OTA aggregation from WDs in distance-$(L-\ell+1)$ set to WDs in distance-$(L-\ell)$ set at slot $\ell$, where $\ell=1,...,L-1$.}\label{fig:level-t-OTA}
    \vspace{-0.5cm}
\end{figure}

 Based on the D2D wireless multihop network topology, as described in Sec.~II-C, we propose a multi-level OTA aggregation based MapReduce scheme to compute a nomographic function at WD $K$ over the multihop D2D network. As illustrated in Fig.~\ref{fig:Tx_Protocol}, we consider a block-based transmission protocol. We examine a quasi-static block fading channel model, in which each D2D communication channel coefficient remains unchanged within one block, but may vary block by block. The block is further divided into $L$ slots of identical duration, and level-$\ell$ OTA aggregation is completed within the $\ell$-th slot in this block.

 To start with, we define sets ${\cal K}_{\text{leaf}}\triangleq \{i\in{\cal K}\setminus\{K\}|~ |{\cal N}_i|=1 \}$ and ${\cal K}_{\rm inter}\triangleq {\cal K}\setminus({\cal K}_{\text{leaf}}\bigcup\{K\})$ as the sets of {\em leaf} WDs and {\em intermediate} WDs, respectively. It is worth emphasizing that each leaf WD $i\in{\cal K}_{\text{leaf}}$ has only one neighboring WD, while each intermediate WD $i\in{\cal K}_{\text{inter}}$ has more than one neighboring WD. There are three categories of WDs in the tree ${\cal T}(K)$, namely, leaf WDs, intermediate WDs, and a root WD $K$. Define
 \begin{align}
 h_{i\rightarrow K} \triangleq 
 \prod_{j=1}^{|{\cal H}({\cal P}_{i\rightarrow K})|} h_{\pi_{j}(i),\pi_{j-1}(i)}, ~~\forall i\in{\cal K}\setminus\{K\},
 \end{align}
 as the {\em effective} channel coefficient from source WD $i$ to destination WD $K$.

\subsubsection{Leaf WD Transmission}
 Each leaf WD $i\in{\cal K}_{\text{leaf}}$ only needs to transmit its partially aggregated IVA $a^{\rm agg}_i$ to its direct parent WD $\pi(i)$ via a D2D communication link. The analog (uncoded) partially aggregated IVA signal to be transmitted by the leaf WD $i$ is expressed as
 \begin{align}\label{eq.x_leaf}
    x_i = \frac{h^*_{i\rightarrow K}}{|h_{i\rightarrow K}|^2} \eta a^{\rm agg}_{i},~~\forall i\in{\cal K}_{\text{leaf}},
 \end{align}
 where $\eta_i\in\mathbb{C}$ denotes the complex-valued transmit coefficient of source WD $i$ to be designed. From \eqref{eq.x_leaf}, it is shown that each leaf WD simply multiplies the (effective) inverse channel coefficient from itself to the destination WD $K$ (i.e., $\frac{h^*_{i\rightarrow K}}{|h_{i\rightarrow K}|^2}$) and multiplies its partially-aggregated-IVA by the transmit coefficient (i.e., $\eta_i$). Accordingly, the leaf WD $i$'s transmit power constraint is
 \begin{align} \label{eq.power-x-leaf}
 |x_i|^2 = \frac{|\eta_i|^2}{|h_{i\rightarrow K}|^2}|a_i^{\rm agg}|^2\leq P_i,~~\forall i\in{\cal K}_{\text{leaf}},
 \end{align}
 where $P_i>0$ denotes the prescribed transmit power budget of WD $i\in{\cal K}_{\text{leaf}}$.

\subsubsection{Intermediate WD Transmission}
 Each intermediate WD $i\in {\cal K}_{\rm inter}$ receives the (aggregated) signal from its neighboring WDs in the lower-distance WD set (i.e., WDs in ${\cal N}_i\setminus\{\pi(i)\}$), and then generates an IVA-embedded symbol by linearly combining the received signal and its local partially aggregated IVA, and finally transmits a signal to its direct parent WD $\pi(i)$ at the next slot. Therefore, the intermediate WD $i$'s transmit signal is written as
 \begin{align}\label{eq.x_intermediate}
     x_i = y_i + \frac{h^*_{i\rightarrow K}}{|h_{i\rightarrow K}|^2} \eta_i a^{\rm agg}_{i},~~\forall i\in{\cal K}_{\text{inter}},
 \end{align}
 where $y_i$ and $\eta_i$ denote the received signal and transmit coefficient of WD $i\in{\cal K}_{\text{inter}}$, respectively.

\subsubsection{Level-$\ell$ OTA Aggregation From Distance-$(L-\ell+1)$ WD Set to Distance-$(L-\ell)$ WD Set}
 As shown in Fig.~\ref{fig:level-t-OTA}, we introduce the level-$\ell$ OTA aggregation from the distance-$(L-\ell+1)$ WD set to the distance-$(L-\ell)$ WD set at slot $\ell$, where $\ell=1,...,L-1$. Based on the relationship of ${\cal K}_{L-\ell+1} = \cup_{j\in{\cal K}_{L-\ell}} {\cal N}_{j}\setminus\{\pi(j)\}$, we decompose the distance-$(L-\ell+1)$ WD set into $|{\cal K}_{L-\ell}|$ subsets, where each subset collects all the WDs in set ${\cal N}_{j}\setminus\{\pi(j)\}$ for $j\in{{\cal K}_{L-\ell}}$. Furthermore, as in Definition~\ref{def:direct-parent} for direct parent WDs, it holds that $\pi(i) = j$, $\forall i\in {\cal N}_{j}\setminus\{\pi(j)\}$. This implies that WD $j$ is the direct parent WD of WD $i\in {\cal N}_{j}\setminus\{\pi(j)\}$. In Fig.~\ref{fig:level-t-OTA}, the direct D2D communications from the WDs in set ${\cal N}_j\setminus\{\pi(j)\}$ to the WD $j\in{\cal K}_{L-\ell}$ form a multi-access channel. 

 To implement the level-$\ell$ OTA aggregation, all the WDs in distance-$(L-\ell+1)$ WD set ${\cal K}_{L-\ell+1}$ transmit signals to their respective direct parent WDs in the distance-$(L-\ell)$ WD set ${\cal K}_{L-\ell}$ over the same resource blocks. As a result, the level-$\ell$ OTA aggregation from the WDs in ${\cal N}_j\setminus\{\pi(j)\}$ to WD $j$ at slot $\ell$ is given as
 \begin{align} \label{eq.level-l-OTA}
    y_j &= \sum_{i\in{\cal N}_j\setminus\{\pi(j)\}} h_{j,i} x_i + n_j,~~\forall j\in{\cal K}_{L-\ell}\cap{\cal K}_{\text{inter}},
 \end{align}
 where $n_j\sim{\cal CN}(0,\sigma_j^2)$ denotes the independent AWGN with zero-mean and variance $\sigma^2$ at WD $j$'s receiver. Note that, in \eqref{eq.level-l-OTA}, each WD $i\in {\cal N}_j\setminus\{\pi(j)\}$ is either a leaf WD or an intermediate WD. Therefore, for all $j\in{\cal K}_{L-\ell}\bigcap{\cal K}_{\text{inter}}$, we have
 \begin{align}\label{eq.x-i}
    x_i = \begin{cases}
    \eqref{eq.x_leaf},&{\text{if}~i\in({\cal N}_j\setminus\{\pi(j)\})\bigcap{\cal K}_{\text{leaf}}}\\
    \eqref{eq.x_intermediate},&{\text{if}~i\in({\cal N}_j\setminus\{\pi(j)\})\bigcap{\cal K}_{\text{inter}}}.
    \end{cases}
 \end{align}
  By substituting $x_i$ in \eqref{eq.x-i} into \eqref{eq.level-l-OTA}, the received signal at WD $j\in{\cal K}_{L-\ell}\bigcap{\cal K}_{\text{inter}}$ is rewritten as
 \begin{align} \label{eq.y-j}
    y_j &= \sum_{i\in{\cal N}_j^+} \frac{h^*_{j\rightarrow K}}{|h_{j\rightarrow K}|^2} \eta_i a_i^{\rm agg} + \sum_{m\in{\cal N}_j^+\cap{\cal K}_{\text{inter}}}h_{m\rightarrow j}n_m + n_j,
 \end{align}
 where the set ${\cal N}_j^+$ collects WD $j$'s neighboring WDs in the lower-distance WD sets (i.e., ${\cal K}_{L-\ell+1},...,{\cal K}_{L}$) via one or more D2D communication hops.

 As a result, by substituting $y_j$ in \eqref{eq.y-j} into \eqref{eq.x_intermediate}, the transmit signal of this WD $j\in{\cal K}_{L-\ell}\cap{\cal K}_{\text{inter}}$ is explicitly expressed as
  \begin{align}\label{eq.x-j}
     x_j = \sum_{i\in{\cal N}_j^+} \frac{h^*_{j\rightarrow K}}{|h_{j\rightarrow K}|^2}\eta a_i^{\rm agg} + \frac{h^*_{j\rightarrow K}}{|h_{j\rightarrow K}|^2} \eta a^{\rm agg}_{j} + \sum_{m\in{\cal N}_j^+\cap{\cal K}_{\text{inter}}}h_{m\rightarrow j}n_m + n_j.
 \end{align}
 Based on \eqref{eq.x-j}, the transmit power constraint of WD $j\in{\cal K}_{L-\ell}\cap{\cal K}_{\text{inter}}$ is given by
 \begin{align}\label{eq.power-x-inter}
 |x_j|^2 = \frac{1}{|h_{j\rightarrow K}|^2}\Bigg|\sum_{i\in{\cal N}_j^+}\eta_i a_i^{\rm agg} + \eta_j a_j^{\rm agg}\Bigg|^2 + \sum_{i\in{\cal N}_j^+\cap{\cal K}_{\text{inter}}}|h_{i\rightarrow j}|^2\sigma_i^2 +\sigma_j^2 \leq P_j,
 \end{align}
 where $P_j>0$ denotes the prescribed transmit power budget of source WD $j\in{\cal K}_{L-\ell}\cap{\cal K}_{\text{inter}}$.

 As such, we complete the procedure of level-$\ell$ OTA aggregation based on \eqref{eq.level-l-OTA} (or equivalently \eqref{eq.y-j}) within this slot $\ell$. By iteratively setting $\ell$ from $\ell=1$ until $\ell=L-1$, we proceed to implement a series of OTA aggregations as in~\eqref{eq.level-l-OTA} from a lower-distance WD set to a higher-distance WD set during the slots $\{1,...,L-1\}$.

 When $\ell=L$, we proceed to implement the level-$L$ OTA aggregation for WDs in the distance-1 set ${\cal K}_1$ to the root WD $K$ at slot $L$. Since the root WD $K$ has no direct parent WD in the MST ${\cal T}(K)$, it is yielded that the set $\{\pi(K)\}$ is an empty set, i.e., $\{\pi(K)\}=\emptyset$. The neighboring WD set ${\cal N}_K$ of WD $K$ is identical to the distance-1 WD set ${\cal K}_1$ in tree ${\cal T}(K)$, i.e., ${\cal N}_K={\cal K}_1$. Based on \eqref{eq.level-l-OTA} and \eqref{eq.y-j}, the level-$L$ OTA aggregation is given as
  \begin{align}\label{eq.L-OTA}
    y_K = \frac{1}{\gamma}\Big(\sum_{i\in{\cal N}^+_K} h_{K,i}x_i + n_K\Big)= \sum_{i=1}^{K-1} \frac{\eta_i}{\gamma}a_i^{\rm agg} + \frac{1}{\gamma}\tilde{n},
 \end{align}
 where $\gamma>0$ denotes the destination WD $K$'s receive filtering factor, the second equality holds from ${\cal N}_K^+ = {\cal K}\setminus\{K\}$, and $\tilde{n}\triangleq \sum_{j\in{\cal K}_{\text{inter}}}h_{j\rightarrow K}n_j + n_K$ collects all the noise terms in the multi-level OTA aggregation procedure. Due to the independence of noise terms $\{n_j\}_{j\in{\cal K}_{\rm inter}}$ and $n_K$, it is thus shown that $\tilde{n} \sim {\cal CN}(0,\sigma^2)$ with $\sigma^2 \triangleq \sum_{j\in{\cal K}_{\text{inter}}} |h_{j\rightarrow K}|^2\sigma^2_j+\sigma_K^2)$.

 \begin{remark}
 Note that our proposed multi-level OTA aggregation scheme only uses $L$ resource blocks to complete the transmission of partially aggregated IVAs $\{a_1^{\rm agg},...,a_{K-1}^{\rm agg}\}$ from the source WDs $\{1,....,K-1\}$ to the destination WD $K$. When compared with the digital communication scheme in Sec.~III-A, which requires ${\cal O}(KQL)$ resource blocks, the proposed analog multi-level OTA aggregation scheme enjoys a superior spectral efficiency. Furthermore, only a common pilot is needed per source WD, and hence, the multi-level OTA aggregation scheme can achieve a substantial reduction in the number of resource blocks as well as pilots.
\end{remark}

\subsubsection{Reconstructing Aggregated-IVA at Destination WD $K$}
 Finally, the destination WD $K$ needs to reconstruct an aggregated IVA version to closely approximate the ground truth $a^{\rm agg}=\sum_{i=1}^K a_i^{\rm agg}$ at the end of the block as the input of the reduce function $\psi$. Denote by $\hat{a}$ the estimate of the aggregated IVA. We consider a linear combiner at WD $K$ with a low complexity, and linearly add WD $K$'s receive signal $y_K$ and its local partially aggregated IVA $a_K^{\rm agg}$. As a result, the estimate $\hat{a}$ at the destination WD $K$ is given as
 \begin{align}\label{eq.a-hat}
 \hat{a} = y_K + a_K^{\rm agg} = \sum_{i=1}^{K-1} \frac{\eta_i}{\gamma}a_i^{\rm agg} + a_K^{\rm agg} + \frac{1}{\gamma}\tilde{n}.
 \end{align}
 
 As shown in \eqref{eq.a-hat}, the reconstructed aggregated-IVA version $\hat{a}$ includes a weighted sum of partially aggregated IVAs $\sum_{i=1}^{K-1}\frac{\eta_i}{\gamma}a_i^{\rm agg}+a_K^{\rm agg}$ and an additive noise $\frac{1}{\gamma}\tilde{n}$. With the obtained $\hat{a}^{\rm agg}$ in \eqref{eq.a-hat}, the destination WD $K$ generates an output $\psi(\hat{a}^{\rm agg})$ as the target nomographic function value. This completes the multi-level OTA aggregation for computing a nomographic function based on MapReduce over this multihop wireless D2D network.

\section{Problem Formulation} \label{section:problem}
In this section, we first derive the MSE expression of the estimated aggregated IVA $\hat{a}$ in \eqref{eq.a-hat}. Then we re-express the source WDs' transmit power constraints in a concise form, and finally present the power-constrained MSE minimization problem.

\subsection{MSE of Aggregated IVA for Reduce Function}
In order to evaluate the approximation performance of the obtained $\psi(\hat{a})$ with respect to the ground truth $\psi(\sum_{i=1}^K a_i^{\rm agg})$, we evaluate the MSE of the aggregated IVA to measure the distortion of the associated nomographic function value.

Let $e\in\mathbb{C}$ denote the computation error between the estimated aggregated IVA $\hat{a}$ and the ground truth $\sum_{i=1}^K{a}_i^{\rm agg}$. With some algebraic manipulations, we are ready to obtain
\begin{align}\label{eq.error}
e = \hat{a} - \sum_{i=1}^K{a}_i^{\rm agg} &= \sum_{i=1}^{K-1} \frac{\eta}{\gamma}a_i^{\rm agg} + a_K^{\rm agg} + \frac{1}{\gamma}\tilde{n} - \sum_{j=1}^{K} a^{\rm agg}_{j} \notag \\
&= \underbrace{\sum_{i=1}^{K-1}\left(\frac{\eta}{\gamma}-1\right)a_i^{\rm agg}}_{\text{ signal-misalignment~induced~error}} +\underbrace{\frac{1}{\gamma}\tilde{n}}_{\text{noise~induced~error}}.
\end{align}
From \eqref{eq.error}, it is shown that the computation error $e$ is of two kinds of errors, namely, signal-misalignment induced error and noise-induced error. Note that the destination WD $K$'s receive filtering factor $\gamma$ plays an important role in balancing the signal-misalignment errors and noise power. By considering the randomization of noise $\tilde{n}$, the computation MSE of the estimate $\hat{a}$ with respect to $\sum_{i=1}^{K-1}a_i^{\rm agg}$ is given by
\begin{align}\label{eq.MSE}
{\rm MSE}(\bm\eta,\gamma) 
 = \mathbb{E}_{\{n_j\}}\Big[ \Big| \sum_{i=1}^{K-1}\left(\frac{\eta}{\gamma}-1\right)a_i^{\rm agg} +\frac{1}{\gamma}\tilde{n} \Big|^2 \Big] =  \left|\frac{1}{\gamma}\bm\eta^T \bm a-\bm 1_{K-1}^T\bm a \right|^2 + \frac{1}{\gamma^2}\sigma^2,
\end{align}
where $\bm \eta \triangleq [\eta_1,...,\eta_{K-1}]^T$, $\bm a\triangleq [a_1^{\rm agg},...,a_{K-1}^{\rm agg}]^T$, $\bm 1_{K-1}$ denotes the all-one vector of dimension $(K-1)$, and the third equality holds from $\tilde{n}\sim{\cal CN}(0,\sigma^2)$ with $\sigma^2 = \sum_{j\in{\cal K}_{\text{inter}}}|h_{j\rightarrow K}|^2\sigma^2_j+\sigma^2_K$.


\subsection{Source WDs' Transmit Power Constraints}
 The transmit power constraints of the source WDs are given in \eqref{eq.power-x-leaf} and \eqref{eq.power-x-inter}. We now express them in a compact form below. Specifically, we construct vector $\bm b_i\in\mathbb{C}^{(K-1)\times 1}$ for $i\in{\cal K}_{\text{leaf}}$ such that the $i$-th entry $[{\bm b}_i]_i=a_i^{\rm agg}$ and the other entries are all zero, i.e., $[{\bm b}_i]_j=0$, $\forall i\neq j$. On the other hand, we construct vector $\bm b_i\in\mathbb{C}^{(K-1)\times 1}$ for $i\in{\cal K}_{\text{inter}}$ by setting the entry $[{\bm b}_i]_j=a_i^{\rm agg}$ if $j\in{\cal N}_i^+\bigcup\{i\}$, and otherwise, $[{\bm b}_i]_j=0$. Furthermore, by taking into account the channel fading and noise effects in multihop transmission from the source WDs to the destination WD $K$, we introduce the {\em effective} power budgets $\{\bar{P}_1,...,\bar{P}_{K-1}\}$ for source WDs as
 \begin{align}\label{eq.bar-power}
 \bar{P}_i \triangleq \begin{cases}
 P_i, &{\text{if}~}i\in{\cal K}_{\text{leaf}},\\
 P_i-\sum_{j\in{\cal N}_i^+\bigcap{\cal K}_{\text{inter}}}|h_{j\rightarrow i}|^2\sigma_j^2-\sigma_i^2, &{\text{if}~}i\in{\cal K}_{\text{inter}}.
 \end{cases}
 \end{align}

 With vectors $\{\bm b_i\}_{i=1}^{K-1}$ and effective power budgets $\{\bar{P}_i\}_{i=1}^{K-1}$ in \eqref{eq.bar-power}, the $(K-1)$ source WDs' transmit power constraints in \eqref{eq.power-x-leaf} and \eqref{eq.power-x-inter} are re-expressed in Lemma~\ref{lem:Tx-power}.

 \begin{lemma}[WDs' Transmit Power Constraints]\label{lem:Tx-power}
 The transmit power constraints of source WDs $\{1,...,K-1\}$ in \eqref{eq.power-x-leaf} and \eqref{eq.power-x-inter} are equivalently expressed in terms of $\{\eta_i\}_{i=1}^{K-1}$ according to
 \begin{align}\label{eq:power-x}
 |\eta_i|^2 \leq \frac{|h_{i\rightarrow K}|^2}{|{\bm 1}^T_{K-1}\bm b_i|^2}\bar{P}_i,~~\forall i\in{\cal K}\setminus\{K\}.
 \end{align}
 \end{lemma}

\subsection{Power-Constrained MSE Minimization Problem}
 In this paper, we pursue an efficient OTA aggregation design for computing nomographic functions, by minimizing the ${\rm MSE}(\bm \eta,\gamma)$ in \eqref{eq.MSE} subject to individual transmit power constraints \eqref{eq:power-x} at the $(K-1)$ source WDs. We jointly optimize the transmit coefficients $\{\eta_j\}$ of the source WDs $\{1,...,K-1\}$ and the receive filtering factor $\gamma$ at destination WD $K$. Mathematically, the power-constrained MSE minimization problem is formulated as
 \begin{subequations}\label{eq.prob1}
 \begin{align}
 ({\cal P}1): &\min_{\bm\eta,\gamma>0}~~
 \left|\frac{1}{\gamma}\bm\eta^T \bm a-\bm 1_{K-1}^T\bm a \right|^2 + \frac{1}{\gamma^2}\sigma^2 \\ 
 &~~{\rm s.t.}~|\eta_i|^2 \leq \frac{|h_{i\rightarrow K}|^2}{|{\bm 1}^T_{K-1}\bm b_i|^2}\bar{P}_i,~~\forall i\in{\cal K}\setminus\{K\}.
 \end{align}
 \end{subequations}
 Due to the coupling of the transmit coefficient vector $\bm \eta$ and receive filtering factor $\gamma$ in the objective function (\ref{eq.prob1}a), problem ($\cal P$1) is a nonconvex optimization problem\cite{Boyd-Book}. For problem ($\cal P$1), in the following Sec.~V, we utilize the primal decomposition technique to obtain its closed-form solution under the special case of a common transmit coefficient at the source WDs (i.e., $\bm \eta =\eta \bm 1_{K-1}$). Subsequently, in Sec.~VI, we obtain the (near-) optimal solution for problem ($\cal P$1) under the general case by employing Dinkelbach's algorithm\cite{Dinkelbach67}.

\section{Optimal Solution of (${\cal P}1$) with $\bm \eta = \eta \bm 1_{K-1}$}
 In this section, we consider a special case of problem (${\cal P}1$) with $\bm \eta = \eta \bm 1_{K-1}$, in which the source WDs in ${\cal K}\setminus\{K\}$ employ a common transmit coefficient. In this case, we employ the primal decomposition approach to analytically solve problem (${\cal P}1$), and analyze the achieved MSE performance under the proposed multi-level OTA aggregation design.

\subsection{Primal Decomposition of (${\cal P}1$) with $\bm \eta = \eta \bm 1_{K-1}$}
 In the case of $\bm \eta = \eta \bm 1_{K-1}$, problem (${\cal P}1$) is reduced as 
 \begin{subequations}\label{eq.prob11}
 \begin{align}
 ({\cal P}1.1):~ &\min_{\eta>0,\gamma>0}~~
\Big(\frac{\eta}{\gamma}-1 \Big)^2|\bm 1_{K-1}^T \bm a|^2+\frac{\sigma^2}{\gamma^2} \\
 &~~{\rm s.t.}~\eta^2 \leq \frac{|h_{j\rightarrow K}|^2}{|{\bm 1}^T_{K-1}\bm b_j|^2}\bar{P}_j,~\forall j\in{\cal K}\setminus\{K\},\label{eq.prob11b}
 \end{align}
 \end{subequations}
 where $\eta\in\mathbb{R}$ is assumed without loss of optimality. Denote by $(\eta^{\rm opt},\gamma^{\rm opt})$ the optimal solution of $({\cal P}1.1)$. Note that the transmit power constraints (\ref{eq.prob11}b) are independent of WD $K$'s receive filtering factor $\gamma$, in the sense that each constraint function of (\ref{eq.prob11}b) depends only on $\eta$. Denote by $\gamma^{\rm opt}(\eta)$ the optimal solution that minimizes ${\rm MSE}(\eta,\gamma)$ under a given $\eta>0$. We then have
 \begin{align}
 ({\cal P}2.1):~~\gamma^{\rm opt}(\eta)\triangleq \argmin_{\gamma>0}~~ {\rm MSE}(\eta,\gamma), \notag
 \end{align}
 where ${\rm MSE}(\eta,\gamma) = \Big(\frac{\eta}{\gamma}-1 \Big)^2|\bm 1_{K-1}^T \bm a|^2+\frac{\sigma^2}{\gamma^2}$. In addition, the power-constrained MSE minimization problem (${\cal P}1.1$) can be equivalently reformulated as
 \begin{align}\label{eq.prob2-2}
 ({\cal P}2.2):~ &\eta^{\rm opt} = \argmin_{\eta>0:~\eqref{eq.prob11b}} ~ {\rm MSE}(\eta,\gamma^{\rm opt}(\eta)). 
 \end{align}
 
 As a result, problem (${\cal P}1$) can be optimally solved by first obtaining the optimal $\gamma^{\rm opt}(\eta)$ that minimizes ${\rm MSE}(\eta,\gamma)$ under a given $\eta$ by problem (${\cal P}2.1$), then minimizing ${\rm MSE}(\eta,\gamma^{\rm opt}(\eta))$ with respect to $\eta$ by problem (${\cal P}2.2$), and finally achieving the optimal $\eta^{\rm opt}$ and $\gamma^{\rm opt}=\gamma^{\rm opt}(\eta^{\rm opt})$.

\subsection{Solutions to (${\cal P}2.1$) and (${\cal P}2.2$)}
 In this subsection, we present the optimal solutions for problems (${\cal P}2.1$) and (${\cal P}2.2$).

\subsubsection{Optimal $\gamma^{\rm opt}(\eta)$ for (${\cal P}2.1$)} First, with algebra manipulations, we recast ${\rm MSE}(\eta,\gamma)$ as
 \begin{align}\label{eq.mse-eta-1}
 &{\rm MSE}(\eta,\gamma) = \frac{\eta^2|\bm 1^T_{K-1} \bm a|^2+\sigma^2}{\gamma^2}- \frac{2\eta |\bm 1_{K-1}^T {\bm a}|^2}{\gamma} + |\bm 1^T_{K-1}\bm a|^2,
 \end{align}
 which is a strictly convex quadratic function of variable $1/{\gamma}$ due to $\eta^2|\bm 1^T_{K-1}\bm a|^2+\sigma^2>0$\cite{Boyd-Book}. In addition, under any given positive transmit coefficient $\eta>0$, minimizing ${\rm MSE}(\eta,\gamma)$ over variable $\gamma$ in \eqref{eq.mse-eta-1} will always lead to a nontrivial solution of $\gamma$. Therefore, problem $({\cal P}2.1)$ is a univariate convex optimization problem. Based on Karush-Kuhn-Tucker (KKT) conditions\cite{Boyd-Book}, we are ready to obtain the optimal solution $\gamma^{\rm opt}(\eta)$ for problem (${\cal P}2.1$) as follows.

 \begin{lemma}[Optimal $\gamma^{\rm opt}(\eta)$ under Given $\eta>0$] \label{lemma:opt-gamma}
 For the given feasible $\eta>0$, the optimal $\gamma^{\rm opt}(\eta)$ to problem (${\cal P}2.1$) is given by
 \begin{align}\label{eq.opt-gamma}
 \gamma^{\rm opt}(\eta) =  \frac{\eta^2|\bm 1^T_{K-1}\bm a|^2 + \sigma^2}{\eta|\bm 1^T_{K-1}\bm a|^2}  =  \eta + \frac{\sigma^2}{\eta|\bm 1^T_{K-1}\bm a|^2}.
 \end{align}
 \end{lemma}

 \begin{IEEEproof}
 See Appendix A.
 \end{IEEEproof}

 \begin{remark}[Asymptotically Unbiasedness of $(\eta,\gamma^{\rm opt}(\eta))$]
 From \eqref{eq.opt-gamma} in Lemma~\ref{lemma:opt-gamma}, it follows that $\frac{\eta}{\gamma^{\rm opt}(\eta)} = \frac{\eta}{\eta + \frac{\sigma^2}{\eta|\bm 1^T_{K-1}\bm a|^2}}<1$. We then have
 \begin{align}
 \mathbb{E}[\hat{a}] = \mathbb{E} \left[\sum_{i=1}^{K-1} \frac{\eta}{\gamma}a_i^{\rm agg} + a_K^{\rm agg} + \frac{1}{\gamma}\tilde{n}\right]=\mathbb{E} \left[\sum_{i=1}^{K-1} \frac{\eta}{\gamma}a_i^{\rm agg} + a_K^{\rm agg} \right]\neq \sum_{i=1}^Ka_i^{\rm agg},
 \end{align}
 which demonstrates that the estimated aggregated IVA $\hat{a}$ at destination WD $K$ is {\em biased}. On the other hand, if the transmit power constraints for source WDs are removed (i.e., the transmit coefficient $\eta$ can be set to be a sufficiently large value), then we have
 \begin{align}
 \lim_{\eta\rightarrow\infty} \frac{\eta}{\gamma^{\rm opt}(\eta)} = 1,
 \end{align}
 which shows that the estimate $\hat{a}$ achieved by $(\eta,\gamma^{\rm opt}(\eta))$ is {\em asymptotically unbiased}.
 \end{remark}

\subsubsection{Obtaining Optimal $\eta^{\rm opt}$ for (${\cal P}2.2$)}

 Next, we proceed to gain the optimal $\eta^{\rm opt}$ for problem (${\cal P}2.2$). By substituting $\gamma^{\rm opt}(\eta)$ in~\eqref{eq.opt-gamma} into ${\rm MSE}(\eta,\gamma)$, the ${\rm MSE}(\eta,\gamma^{\rm opt}(\eta))$ is expressed as
 \begin{align}\label{eq.mse-eta}
 {\rm MSE}(\eta,\gamma^{\rm opt}(\eta)) =
 |\bm 1^T_{K-1}\bm a|^2- \frac{\eta^2|\bm 1^T_{K-1}\bm a|^4}{\eta^2|\bm 1^T_{K-1}\bm a|^2+\sigma^2} = \frac{\sigma^2}{\eta^2 + \sigma^2/|\bm 1^T_{K-1}\bm a|^2}.
 \end{align}

Based on \eqref{eq.mse-eta}, we can minimize ${\rm MSE}(\eta,\gamma^{\rm opt}(\eta))$ by equivalently maximizing the magnitude of transmit coefficient $\eta$. With the transmit power constraints in problem (${\cal P}2.2$), we then have
\begin{align}
 \eta \leq \min_{i=1,...,K-1} \frac{|h_{i\rightarrow K}|}{|\bm 1^T_{K-1} \bm b_i|} \sqrt{\bar{P}_i}.
\end{align}
The optimal solution for problem (${\cal P}2.2$) is $(\eta^{\rm opt},\gamma^{\rm opt})$ obtained as follows.

\begin{proposition}[Optimal Transmit Coefficient $\eta^{\rm opt}$]\label{prop:eta-opt}
 At the optimality of problem (${\cal P}2.2$) (and, equivalently, (${\cal P}1.1$)), the source WDs' transmit coefficient $\eta^{\rm opt}$ is given by
\begin{align}
\eta^{\rm opt} &= \min_{i=1,...,K-1} \frac{|h_{i\rightarrow K}|}{|\bm 1^T_{K-1} \bm b_i|}\sqrt{\bar{P}_i}.
\end{align}
\end{proposition}

\begin{proposition}[Optimal Receive Filtering Factor $\gamma^{\rm opt}$]\label{prop:gamma-opt}
By substituting $\eta^{\rm opt}$ into \eqref{eq.opt-gamma} in Lemma~\ref{lemma:opt-gamma}, the optimal receive filtering factor $\eta^{\rm opt}$ is obtained as
\begin{align}
\gamma^{\rm opt} = \gamma^{\rm opt}(\eta^{\rm opt}) = \eta^{\rm opt} + \frac{\sigma^2}{ \eta^{\rm opt} |\bm 1^T_{K-1}\bm a|^2 }.
\end{align}
\end{proposition}

Based on Propositions \ref{prop:eta-opt} and \ref{prop:gamma-opt}, we finally obtain the optimal solution $(\eta^{\rm opt},\gamma^{\rm opt})$ of the power-constrained MSE minimization problem (${\cal P}1.1$). 


\subsection{MSE Performance Analysis}
 As previously mentioned, the estimate $\hat{a}$ achieved by the optimal $(\eta^{\rm opt},\gamma^{\rm opt})$ is biased, but asymptotically unbiased. With $(\eta^{\rm opt},\gamma^{\rm opt})$ obtained, the ${\rm MSE}(\eta^{\rm opt},\gamma^{\rm opt})$ achieved by our proposed multi-level OTA aggregation scheme is given by
 \begin{align}\label{eq.mse-biased-opt}
 {\rm MSE}(\eta^{\rm opt},\gamma^{\rm opt}) 
 =\frac{\sigma^2 }{P_{\min} + \sigma^2/|\bm 1^T_{K-1}\bm a|^2}, 
 \end{align}
 where $(\eta^{\rm opt})^2=P_{\min} \triangleq \min_{i=1,...,K-1} \frac{|h_{i\rightarrow K}|^2  }{|\bm1^T_{K-1} \bm b_i|^2}\bar{P}_i$ is defined as the minimal ratio of the achievable power gain to the transmit signal power among all the source WDs $ \{1,...,K-1\}$.

 For performance comparison, we consider an unbiased design scheme such that $\mathbb{E}[\hat{a}]=\sum_{i=1}^K a_i^{\rm agg}$. This can be realized by setting $\gamma=\eta$; i.e., the source WDs' transmit coefficient $\eta$ is identical to the destination WD's receive filtering factor $\gamma$. In this case, as shown in \eqref{eq.error}, the estimation error includes only the noise-induced error (i.e., $\frac{1}{\eta}\tilde{n}$). This implies that the signal-misalignment-induced error is forced to be zero in this unbiased scheme. Therefore, minimizing the MSE is equivalent to minimizing the noise variance for $\frac{1}{\eta}\tilde{n}$. Accordingly, the minimal MSE achieved by this unbiased estimate $\hat{a}$ is obtained as
 \begin{align}\label{eq.mse-unbiased}
 {\rm MSE}_{\text{unbiased}} &= \min_{\eta^2\leq P_{\min}} \frac{\sigma^2}{\eta^2} = \frac{\sigma^2}{P_{\min}}, 
 \end{align}
 and the optimal transmit coefficient and receive filtering factor are $\gamma_{\text{unbiased}} = \eta_{\text{unbiased}}=\sqrt{P_{\min}}$.

 Compared to the unbiased scheme enforcing $\eta=\gamma$, our proposed (biased) scheme admits a larger degree of freedom in system design since it removes the constraint of $\eta = \gamma$. This in turn enables the proposed scheme to achieve a smaller MSE value under the source WDs' transmit power constraints. More specifically, we have
 \begin{align}\label{eq.comp}
 \begin{cases}
 \eta^{\rm opt} = \eta_{\text{unbiased}}=\sqrt{P_{\min}}\\
 \gamma^{\rm opt} > \gamma_{\text{unbiased}}.
 \end{cases}
 \end{align}
From the ${\rm MSE}(\bm\eta,\gamma)$ in \eqref{eq.MSE}, it follows that a larger value of the receive filtering factor $\gamma$ leads to a better performance in suppressing the noise-induced error. Due to $\gamma^{\rm opt}>\gamma_{\text{unbiased}}$ in \eqref{eq.comp}, the proposed (biased) scheme is more effective in suppressing the noise-induced error at the expense of allowing some signal-misalignment-induced errors. Therefore, to minimize the MSE of the aggregated IVA, it is important for multi-level OTA aggregation design to strike a balance between minimizing signal-misalignment-induced errors and minimizing noise-induced error.

\begin{figure}
  \centering
  \includegraphics[width = 3.5in]{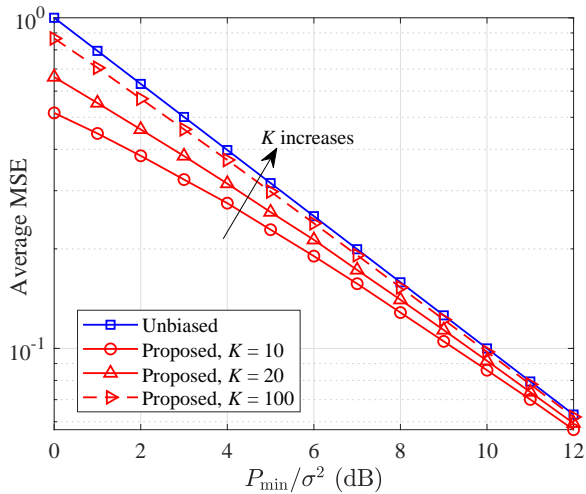}
  \vspace{-0.5cm}
  \caption{The average MSE performance versus $P_{\min}/\sigma^2$, where the partially-aggregated-IVAs are $a_i^{\rm agg}\sim{\cal CN}(0,0.2)$, $\forall i\in{\cal K}\setminus\{K\}$.} \label{fig.MSE_vs_SNR}
  \vspace{-0.5cm}
 \end{figure}

 Fig.~\ref{fig.MSE_vs_SNR} illustrates the MSE performance versus the ratio of $P_{\min}/\sigma^2$ for the proposed joint design scheme \eqref{eq.mse-biased-opt} and the unbiased one \eqref{eq.mse-unbiased}. It is expected that the MSE value should decrease as $P_{\min}/\sigma^2$ increases. At a low regime of $P_{\min}/\sigma^2$, the proposed scheme achieves a significant gain in MSE, but this gain decreases with the increase of $P_{\min}/\sigma^2$. In addition, the achieved MSE gain decreases as the number of WDs $K$ increases. This is because a large number of WDs implies a severe effect from signal-misalignment error, and the proposed scheme tends to suppress the signal-misalignment error to minimize the MSE of interest.

\vspace{-0.4cm}
\section{General Solution to (${\cal P}1$) based on Dinkelbach's Algorithm}
In this section, we use Dinkelbach's algorithm to obtain the (near-) optimal solution of problem (${\cal P}1$) with independent transmit coefficients for the source WDs.

\subsection{Problem Transformation}
By employing primal decomposition for problem (${\cal P}1$), the destination WD $K$'s optimal receive filtering factor $\gamma^{\rm opt}(\bm \eta)$ under the given $\bm \eta$ is obtained as
\begin{align}\label{eq.opt-gamma-vec}
    \gamma^{\rm opt}(\bm \eta) \triangleq \Big(\argmin_{\gamma>0}~~ {\rm MSE}(\bm \eta,\gamma) \Big) = \frac{|\bm \eta^T\bm a|^2 + \sigma^2}{{\rm Re}(\bm\eta^T \bm a \bm 1^T_{K-1}\bm a)}.
\end{align}

Substituting \eqref{eq.opt-gamma-vec} into ${\rm MSE}(\bm \eta,\gamma)$ in \eqref{eq.MSE}, it is yielded that
\begin{align}\label{eq.mse-vec}
    {\rm MSE}(\bm \eta,\gamma^{\rm opt}(\bm\eta)) &= |\bm 1_{K-1}^T \bm a|^2 - \frac{{\rm Re}^2(\bm\eta^T \bm a \bm 1^T_{K-1}\bm a)}{|\bm \eta^T \bm a|^2 + \sigma^2} =|\bm 1_{K-1}^T\bm a|^2\times \frac{\sigma^2 + {\rm Im}^2(e^{j\theta}\bm\eta^T\bm a)}{|\bm \eta^T \bm a|^2 + \sigma^2},
\end{align}
where $e^{j\theta}=\bm 1_{K-1}^T\bm a/|\bm 1_{K-1}^T\bm a|$ with $j=\sqrt{-1}$ being the imaginary unit. For the ${\rm MSE}(\bm \eta,\gamma^{\rm opt}(\bm\eta))$ in \eqref{eq.mse-vec}, we establish the following lemma.
\begin{lemma}[Monotonically Decreasing ${\rm MSE}(\bm \eta,\gamma^{\rm opt}(\bm\eta))$ in $\|\bm \eta\|^2$]\label{lem.mono-mse}.
The ${\rm MSE}(\bm \eta,\gamma^{\rm opt}(\bm\eta))$ is a monotonically decreasing function with respect to $\|\bm \eta\|^2$.
\end{lemma}
\begin{IEEEproof}
See Appendix~B.
\end{IEEEproof}

Lemma~\ref{lem.mono-mse} indicates that, under a given profile $\tilde{\bm \eta} =\bm \eta/\|\bm \eta\|$ of the unit-norm transmit coefficient vector, one can always employ more transmit power to decrease the computation distortion. This is intuitively expected, since a larger transmit power with a fixed $\tilde{\bm \eta}$ leads to a higher SNR in constructing a copy of the aggregated IVA $\sum_{i=1}^{K-1} a_i^{\rm agg}$ at WD $K$.

Furthermore, we simplify the ${\rm MSE}(\bm \eta,\gamma^{\rm opt}(\bm\eta))$ by removing the imaginary operation in \eqref{eq.mse-vec}. First, we define ${\bm \eta}_{\rm ext} \triangleq [{\rm Re}({\bm \eta})^T, {\rm Im}({\bm \eta})^T]^T$, $\bm a_1 \triangleq [{\rm Re}^T(\bm a),-{\rm Im}^T(\bm a)]^T $, $\bm a_2 \triangleq [{\rm Im}^T(\bm a),{\rm Re}^T(\bm a)]^T$, and $\bm a_3 \triangleq [{\rm Im}^T(e^{j\theta}\bm a),{\rm Re}^T(e^{j\theta} \bm a)]^T$. Then, the ${\rm MSE}(\bm \eta,\gamma^{\rm opt}(\bm\eta))$ in \eqref{eq.mse-vec} is rewritten in terms of ${\bm \eta}_{\rm ext}$ as
 \begin{align}\label{eq:mse-opt-vector}
    {\rm MSE}(\bm \eta,\gamma^{\rm opt}(\bm\eta)) = |\bm 1_{K-1}^T\bm a|^2 \times \frac{ {\bm\eta}_{\rm ext}^T\bm a_3\bm a_3^T {\bm \eta}_{\rm ext} +{\sigma^2} }{ {\bm\eta}^T_{\rm ext}(\bm a_1\bm a_1^T + \bm a_2 \bm a_2^T) {\bm \eta}_{\rm ext} + {\sigma^2} }. 
 \end{align}

 Regarding the individual transmit power constraint in (\ref{eq.prob1}b) for the source WDs $\{1,...,K-1\}$, we re-express them as
 \begin{align}\label{eq.tx-p}
    [\bm\eta_{\rm ext}]_i^2 + [\bm\eta_{\rm ext}]_{i+K-1}^2  \leq \frac{|h_{i\rightarrow K}|^2}{|{\bm 1}^T_{K-1}\bm b_i|^2} \bar{P}_i,~\forall i\in{\cal K}\setminus\{K\}.
 \end{align}
 As a result, by removing the irrelevant term $|\bm 1_{K-1}^T\bm a|^2$ with respect to $\bm \eta_{\rm ext}$ from ${\rm MSE}(\bm \eta,\gamma^{\rm opt}(\bm\eta))$ in \eqref{eq:mse-opt-vector}, (${\cal P}1$) can be formulated as a constrained quadratic fractional programming problem\cite{Dinkelbach67}:
 \begin{align}\label{eq.prob3}
 ({\cal P}3):~~ &\xi^\star = \min_{{\bm\eta}_{\rm ext}\in{\cal B}}~~
\frac{{\bm\eta}_{\rm ext}^T\bm a_3\bm a_3^T {\bm \eta}_{\rm ext} +{\sigma^2} }{ {\bm\eta}^T_{\rm ext}(\bm a_1\bm a_1^T + \bm a_2 \bm a_2^T) {\bm \eta}_{\rm ext} + {\sigma^2} },
  \end{align}
 where ${\cal B}\triangleq\left\{\bm \eta \big|[\bm\eta_{\rm ext}]_i^2 + [\bm\eta_{\rm ext}]_{i+K-1}^2  \leq \frac{|h_{i\rightarrow K}|^2}{|{\bm 1}^T_{K-1}\bm b_i|^2} \bar{P}_i, \forall i\in{\cal K}\setminus\{K\}\right\}$ deotes the constraint set by \eqref{eq.tx-p}, and the optimal value of problem (${\cal P}3$) is defined as $\xi^\star$. Note that problem (${\cal P}3$) involves minimizing the ratio of two quadratic functions subject to $(K-1)$ quadratic inequality constraints, which is challenging to solve directly due to its nonconvexity\cite{Boyd-Book}. 
 
 In order to handle such nonconvexity, we define a function $ F(\bm \eta_{\rm ext},\xi)$ which is associated with the objective function of $({\cal P}3)$ as
\begin{align}\label{eq.F-func}
    F(\bm \eta_{\rm ext},\xi) \triangleq {\bm\eta}_{\rm ext}^T\bm a_3\bm a_3^T {\bm \eta}_{\rm ext} +{\sigma^2} - \xi( {\bm\eta}^T_{\rm ext}(\bm a_1\bm a_1^T + \bm a_2 \bm a_2^T) {\bm \eta}_{\rm ext} + {\sigma^2} ),
\end{align}
where $\xi\geq 0$ is the auxiliary variable.

\begin{proposition}[Optimal Condition for $\xi^\star$]\label{prop:opt-xi}
Denote by $\bm \eta^\star_{\rm ext}$ the optimal solution of (${\cal P}3$). The optimal $\xi^\star$ is attained such that  
\begin{align}
\left(\min_{\bm \eta_{\rm ext}\in{\cal B}}  F(\bm \eta_{\rm ext},\xi^\star) \right) =  F(\bm \eta^\star_{\rm ext},\xi^\star) = 0.
\end{align}
\end{proposition}
\begin{IEEEproof}
See Appendix C.
\end{IEEEproof} 

Proposition~\ref{prop:opt-xi} states that one can optimally solve problem $({\cal P}3)$ by finding the root of the univarate equation $\left(\min_{\bm \eta_{\rm ext}\in{\cal B}}  F(\bm \eta_{\rm ext},\xi) \right)=0$ with respect to $\xi>0$. In the next subsection, we apply Dinkelbach's algorithm to solve $\min_{\bm \eta_{\rm ext}\in{\cal B}}  F(\bm \eta_{\rm ext},\xi)$ with an updated $\xi$ value per iteration, towards achieving the (near-) optimal solution of (${\cal P}3$) with guaranteed convergence.

\subsection{Solving (${\cal P}3$) based on Dinkelbach's Algorithm}
In this subsection, we apply Dinkelbach's algorithm to (near-) optimally solve problem (${\cal P}3$). The basic idea of Dinkelbach's algorithm is to transform the original fractional programming problem into a series of parametric problems by introducing an auxiliary variable\cite{Dinkelbach67}. These parametric problems with the updated auxiliary variable are iteratively solved until convergence.

Based on this idea, in the $t$th iteration, we need to solve the quadratic programming problem as
\begin{align}\label{eq.def1-eta-t}
\bm \eta_{\rm ext}[t+1] \triangleq \argmin_{\bm \eta_{\rm ext}\in{\cal B}} F(\bm \eta_{\rm ext},\xi[t]),
\end{align}
where $\bm \eta_{\rm ext}[t+1]$ denotes the optimal solution of problem \eqref{eq.def1-eta-t}, and the auxiliary variable $\xi[t]$ is updated by
\begin{align} \label{eq.xi-t}
    \xi[t] = \frac{{\bm\eta}_{\rm ext}^T[t]\bm a_3\bm a_3^T {\bm \eta}_{\rm ext}[t] +{\sigma^2}}{{\bm\eta}^T_{\rm ext}[t](\bm a_1\bm a_1^T + \bm a_2 \bm a_2^T) {\bm \eta}_{\rm ext}[t] + {\sigma^2}},
\end{align}
with $t$ being the iteration index. Based on \eqref{eq.def1-eta-t} and \eqref{eq.xi-t}, we have 
\begin{align*}
    F(\bm \eta_{\rm ext}[t+1],\xi[t]) = \min_{\bm \eta_{\rm ext}\in{\cal B}} F(\bm \eta_{\rm ext},\xi[t]) \leq F(\bm \eta_{\rm ext}[t],\xi[t])=0,
\end{align*}
which further leads to  
\begin{align}\label{eq.decreasing}
 \xi[t+1] = \frac{{\bm\eta}_{\rm ext}^T[t]\bm a_3\bm a_3^T {\bm \eta}_{\rm ext}[t] +{\sigma^2}}{{\bm\eta}^T_{\rm ext}[t](\bm a_1\bm a_1^T + \bm a_2 \bm a_2^T) {\bm \eta}_{\rm ext}[t+1] + {\sigma^2}} \leq \xi[t].
\end{align}
From \eqref{eq.decreasing}, it follows that the sequence $\{\xi[t]\}$ is non-increasing after each iteration. Indeed, due to the monotonicity of $\{\xi[t]\}$, the convergence to an optimal solution of problem $({\cal P}3)$ by Dinkelbach's algorithm is guaranteed via iteratively updating $\xi$ according to \eqref{eq.xi-t} and solving \eqref{eq.def1-eta-t} for $\bm \eta_{\rm ext}$. The iteration by the Dinkelbach's algorithm is shown to converge at a superlinear convergence rate\cite{Schiable76,Yu18-TSP}. We denote by $\bm \eta_{\rm ext}^\star$ the solution achieved by the algorithm.

\begin{algorithm}
\caption{\em for Solving MSE Minimization Problem $({\cal P}1)$ based on Dinkelbach's Algorithm}\label{alg2}
\begin{algorithmic}[1]
\State 
{\bf Initialization:}
Given the partially aggregated IVAs $\{ a^{\rm agg}_i\}_{i=1}^{K-1}$ of the source WDs, a number of D2D multihop effective channel coefficients $\{h_{i\rightarrow K}\}$ from WD $i$ to WD $K$, and WD $j$'s receiver AWGN variance $\sigma_j^2$, $\forall j\in{\cal K}_{\text{inter}}\cup\{K\}$; set $\epsilon>0$ as the convergence tolerance; set iteration index $t=0$ and initialize $\bm \eta_{\rm ext}[0]$;
 \State 
 {\bf Set} $\bm a_1\gets [{\rm Re}^T(\bm a),-{\rm Im}^T(\bm a)]^T$, $\bm a_2\gets [{\rm Im}^T(\bm a),{\rm Re}^T(\bm a)]^T$, $\bm a_3\gets [{\rm Im}^T(e^{j\theta}\bm a),{\rm Re}^T(e^{j\theta}\bm a)]^T$, where $e^{j\theta} \gets \bm 1_{K-1}^T\bm a/|\bm 1_{K-1}^T\bm a|$;
 \State
 {\bf Repeat:} 
\begin{itemize}
  \item Update $\xi[t] \gets \frac{{\bm\eta}_{\rm ext}^T[t]\bm a_3\bm a_3^T {\bm \eta}_{\rm ext}[t] +{\sigma^2}}{{\bm\eta}^T_{\rm ext}[t](\bm a_1\bm a_1^T + \bm a_2 \bm a_2^T) {\bm \eta}_{\rm ext}[t] + {\sigma^2}}$ according to \eqref{eq.xi-t};
   \item Obtain $\bm X[t+1]$ by solving the semidefinite programming problem \eqref{eq.SDR-prob};
    \item {\bf if} $\{{\rm rank}(\bm X[t+1])=1\}$, {\bf then} 
    \item[] \quad Obtain the optimal $\bm \eta[t+1]$ of \eqref{eq.prob-Bt} based on the EVD for $\bm X[t+1]$;  \item[] {\bf else} $\{{\rm rank}(\bm X[t+1])=1 \}$
     \item[] \quad Obtain the near-optimal $\bm \eta[t+1]$ of \eqref{eq.prob-Bt} based on Gaussian randomization method~\cite{Luo10-SPM};\\
    {\bf endif}
    \item Update $t\gets t+1$;
\end{itemize}
 {\bf Until} Convergence condition is satisfied with $\xi[t-1] - \xi[t] <\epsilon$.
 \State 
{\bf Set} $\bm \eta^\star_{\rm ext} \gets \bm \eta_{\rm ext}[t+1]$ as the (near-) optimal solution $\bm \eta^\star_{\rm ext}$ of problem (${\cal P}3$);
\State
{\bf Set} $[\bar{\bm \eta}]_i \gets [\bm \eta^\star_{\rm ext}]_{i} + \sqrt{-1}\times [\bm \eta^\star_{\rm ext}]_{i+K-1}$, $\forall i\in{\cal K}\setminus\{K\}$ by~\eqref{eq.bar-eta};
\State
{\bf Set} $\bar{\gamma}\gets \frac{|\bar{\bm \eta}^T\bm a|^2 + \sigma^2}{{\rm Re}(\bar{\bm\eta}^T \bm a \bm 1^T_{K-1}\bm a)}$ by \eqref{eq.opt-gamma-vec}; 
\State
{\bf Output}: Obtain the (near-) optimal solution $(\bar{\bm \eta},\bar{\gamma})$ of (${\cal P}1$).
\end{algorithmic}
\end{algorithm}

\subsubsection{Obtaining $\bm \eta_{\rm ext}[t+1]$ by \eqref{eq.def1-eta-t} for $\xi[t]$} As previously mentioned, the $t$th iteration of Dinkelbach's algorithm involves solving problem \eqref{eq.def1-eta-t}. By introducing $\bm B(\xi[t])\triangleq \bm a_3\bm a_3^T - \xi[t](\bm a_1\bm a_1^T + \bm a_2 \bm a_2^T)$, problem~\eqref{eq.def1-eta-t} is rewritten as
\begin{align}\label{eq.prob-Bt}
    \bm \eta_{\rm ext}[t+1] &=\argmin_{\bm \eta_{\rm ext}\in{\cal B}} \bm \eta_{\rm ext}^T \bm B(\xi[t]) \bm \eta_{\rm ext}+(1-\xi[t])\sigma^2 \notag \\
    &=\argmin_{\bm \eta_{\rm ext}\in{\cal B}} \bm \eta_{\rm ext}^T \bm B(\xi[t]) \bm \eta_{\rm ext}.
\end{align}
 Since $\bm B(\xi[t])$ has both positive and negative eigenvalues $\{\|\bm a_3\|^2,-\xi[t]\|\bm a_1\|^2,-\xi[t]\|\bm a_2\|^2\}$ under the independence of vectors $\{\bm a_1,\bm a_2,\bm a_3\}$, the matrix $\bm B(\xi[t])$ is generally an indefinite matrix. Hence, problem \eqref{eq.prob-Bt} involves minimizing a non-convex quadratic function over the convex constraint set $\cal B$ for vector $\bm \eta_{\rm ext}$, which is a non-convex optimization problem. We next employ the semidefinite relaxation (SDP) approach to solve \eqref{eq.prob-Bt}.

 Define $\bm X \triangleq \bm \eta_{\rm ext}\bm \eta_{\rm ext}^T\in\mathbb{R}^{2(K-1)\times 2(K-1)}$. It always holds that $\bm X \succeq \bm 0$ and ${\rm rank}(\bm X)=1$. Removing the rank-1 constraint, \eqref{eq.prob-Bt} is relaxed into a semidefinite programming (SDP) problem:
 \begin{align}\label{eq.SDR-prob}
 \bm X[t+1]=\argmin_{\bm X\succeq \bm 0} &~ {\rm tr}(\bm B(\xi[t])\bm X) \notag \\
{\rm s.t.}&~~ [\bm X]_{i,i} + [\bm X]_{i+K-1,i+K-1} \leq \frac{|h_{i\rightarrow K}|^2}{|{\bm 1}^T_{K-1}\bm b_i|^2} \bar{P}_i,~\forall i\in{\cal K}\setminus\{K\},
\end{align}
where $\bm X[t+1]$ is the optimal solution to problem \eqref{eq.SDR-prob}. Using off-the-shelf convex solvers (e.g., CVX toolbox \cite{Boyd-Book}), one can efficiently obtain its optimal solution $\bm X[t+1]$. Due to the absence of constraint ${\rm rank}(\bm X)=1$, the matrix $\bm X[t+1]$ is not generally guaranteed to be of rank-1. Based on $\bm X[t+1]$, we proceed to recover the (near-) optimal solution to problem \eqref{eq.prob-Bt} as follows.
\begin{itemize}
    \item If ${\rm rank}(\bm X[t+1])=1$, then we obtain the optimal $\bm \eta_{\rm ext}[t+1]$ of \eqref{eq.prob-Bt} by implementing the eigenvalue decomposition (EVD) such that $\bm X[t+1]=\bm \eta_{\rm ext}[t+1] \bm \eta_{\rm ext}[t+1]$;
    \item If ${\rm rank}(\bm X[t+1] ) > 1$, then we obtain a near-optimal $\bm \eta_{\rm ext}[t+1]$ of \eqref{eq.prob-Bt} by using a Gaussian randomization method\cite{Luo10-SPM}. Given a number of randomization rounds $M$, for $m=1,...,M$, we generate a vector $\bm x_m\sim {\cal N}(\bm 0,\bm X[t+1])$ and construct a feasible solution $d_m \bm x_m$, where $d_m=\argmin_{d\bm x_m\in\mathbb{R}:\bm \eta_{\rm ext}\in{\cal B}} \bm \eta_{\rm ext}^T \bm B(\xi[t]) \bm \eta_{\rm ext}$. Then, we determine the best one $d_{m^\star}\bm x_{m^\star}$ among the feasible solutions $\{d_m\bm x_m\}_{m=1}^M$, where $m^\star=\argmin_{m=1,...,M} F(d_m\bm x_m,\xi[t])$. Finally, we output $\bm \eta_{\rm ext}[t+1] = d_{m^\star} \bm x_{m^\star}$ as the near-optimal solution of \eqref{eq.prob-Bt} under a given $\xi[t]$.
\end{itemize}

 \subsection{Obtaining (near-) Optimal Solution of (${\cal P}1$)}
 With the $\bm \eta_{\rm ext}^\star$ obtained based on Dinkelbach's algorithm, we proceed to construct a convex-valued vector $\bar{\bm \eta}$ for problem (${\cal P}1$) by setting the $i$-th entry as
  \begin{align} \label{eq.bar-eta}
     [\bar{\bm \eta}]_i = [\bm \eta_{\rm ext}^\star]_{i} + \sqrt{-1}\times [\bm \eta^\star_{\rm ext}]_{i+K-1},~~\forall i\in{\cal K}\setminus\{K\}.
 \end{align}
 By replacing $\bm \eta$ with $\bar{\bm \eta}$ in \eqref{eq.opt-gamma-vec}, the destination WD $K$'s optimal receive filtering factor of problem (${\cal P}1$) is obtained as $\bar{\gamma} = \gamma^{\rm opt}(\bar{\bm \eta})=\frac{|\bar{\bm \eta}^T\bm a|^2 + \sigma^2}{{\rm Re}(\bar{\bm\eta}^T \bm a \bm 1^T_{K-1}\bm a)}$. 

 Until now, we finally obtain the (near-) optimal solution $(\bar{\bm \eta}, \bar{\gamma})$ for problem (${\cal P}1$). In summary, we present Algorithm~\ref{alg2} to (approximately) solve problem $({\cal P}1)$.

\section{Numerical Results}
 In this section, numerical results are provided to gauge the performance of the proposed multi-level OTA aggregation design with joint transmit and receive optimization. For an estimated version $\hat{x}$ of the ground truth $x$, we consider the normalized MSE, $\text{Normalized-MSE} = \frac{\mathbb{E}[|x-\hat{x}|^2]}{\mathbb{E}[|x|^2]}$, as a measure of the MSE performance. The simulation results are obtained averaging 1000 wireless D2D Rayleigh fading channel and IVA realizations. For performance comparison, we consider the following three baseline schemes.
 \begin{itemize}
  \item {\em OTA aggregation scheme with individual transmit coefficients based on Rayleigh quotient minimization method:} In this scheme, the source WDs' individual transmit coefficients $\{\eta_i\}$ are obtained by transforming the ${\rm MSE}(\bm \eta,\gamma^{\rm opt}(\bm \eta))$ in \eqref{eq:mse-opt-vector} to a Rayleigh quotient expression. Denote the solution of this scheme by $(\tilde{\bm \eta}_{\text{Ray}},\tilde{\gamma}_{\text{Ray}})$. Specifically, by setting $\|\bm \eta\|^2 = (K-1)P_{\min}$, we need to solve the Rayleigh quotient minimization problem as
  \begin{align} 
  \tilde{\bm\eta}^{\star}_{\rm ext} =\argmin_{\tilde{\bm\eta}_{\rm ext}:\|\tilde{\bm\eta}_{\rm ext}\|=1}~~
  \frac{\tilde{\bm\eta}_{\rm ext}^T \left(\bm a_3\bm a_3^T  + \frac{\sigma^2}{\|{\bm \eta}\|^2} \bm I_{2(K-1)}\right) \tilde{\bm \eta}_{\rm ext} } {\tilde{\bm\eta}_{\rm ext}^T\left(\bm a_1\bm a_1^T + \bm a_2 \bm a_2^T + \frac{\sigma^2}{\|{\bm \eta}\|^2} \bm I_{2(K-1)} \right) \tilde{\bm \eta}_{\rm ext} },
  \end{align}
  where $\tilde{\bm\eta}^{\star}_{\rm ext}$ is obtained as the unit-norm eigenvector associated with the smallest eigenvalue of matrix $\Big(\bm a_1\bm a_1^T + \bm a_2 \bm a_2^T + \frac{\sigma^2}{\|{\bm \eta}\|^2} \bm I_{2(K-1)} \Big)^{-1} \Big(\bm a_3\bm a_3^T  +\frac{\sigma^2}{\|{\bm \eta}\|^2} \bm I_{2(K-1)} \Big)$. By replacing $\bm \eta_{\rm ext}$ with $\tilde{\bm \eta}^\star_{\rm ext}$ in \eqref{eq.bar-eta}, a unit-norm complex-valued transmit coefficient vector is obtained as $\tilde{\bm \eta}$. By guaranteeing each transmit power constraint is not violated, we now attain the source WDs' transmit coefficient vector as $\tilde{\bm \eta}_{\text{Ray}} = \beta \tilde{\bm \eta}$, where $\beta=\min_{i=1,...,K-1}{\bar{P}_i}/{\tilde{\bm \eta}}$. Then, the destination WD $K$'s receive filtering factor is obtained as $\tilde{\gamma}_{\text{Ray}}=\gamma^{\rm opt}(\tilde{\bm \eta})$. 

 \item {\em OTA aggregation scheme with a common transmit coefficient \cite{Zhu19-IOTJ}:} In this scheme, all the $(K-1)$ WDs adopt the same transmit coefficient, i.e., $\eta_i=\eta$. The optimal solution is presented in Propositions 1 and 2, and the MSE performance is also analyzed in Sec.~V--C. 
 
 \item {\em Digital uncoded QAM transmission scheme with $Q$-bit quantization \cite{Xiao06-TSP}:} In this scheme, each source WD $i\in{\cal K}\setminus\{K\}$ first employs a uniform randomized quantizer to generate a unbiased messages, and then adopts a $2^Q$-QAM transmission scheme \cite{Cui05-TWC} for the $Q$ quantized bits. In the simulations, we consider uncoded 4-QAM and 16-QAM digital transmissions by setting $Q=2$ and $Q=4$, respectively.
 \end{itemize}

 \begin{figure*} 
\centering
\begin{minipage}[t]{0.48\textwidth}
\centering
\includegraphics[width=3.4in]{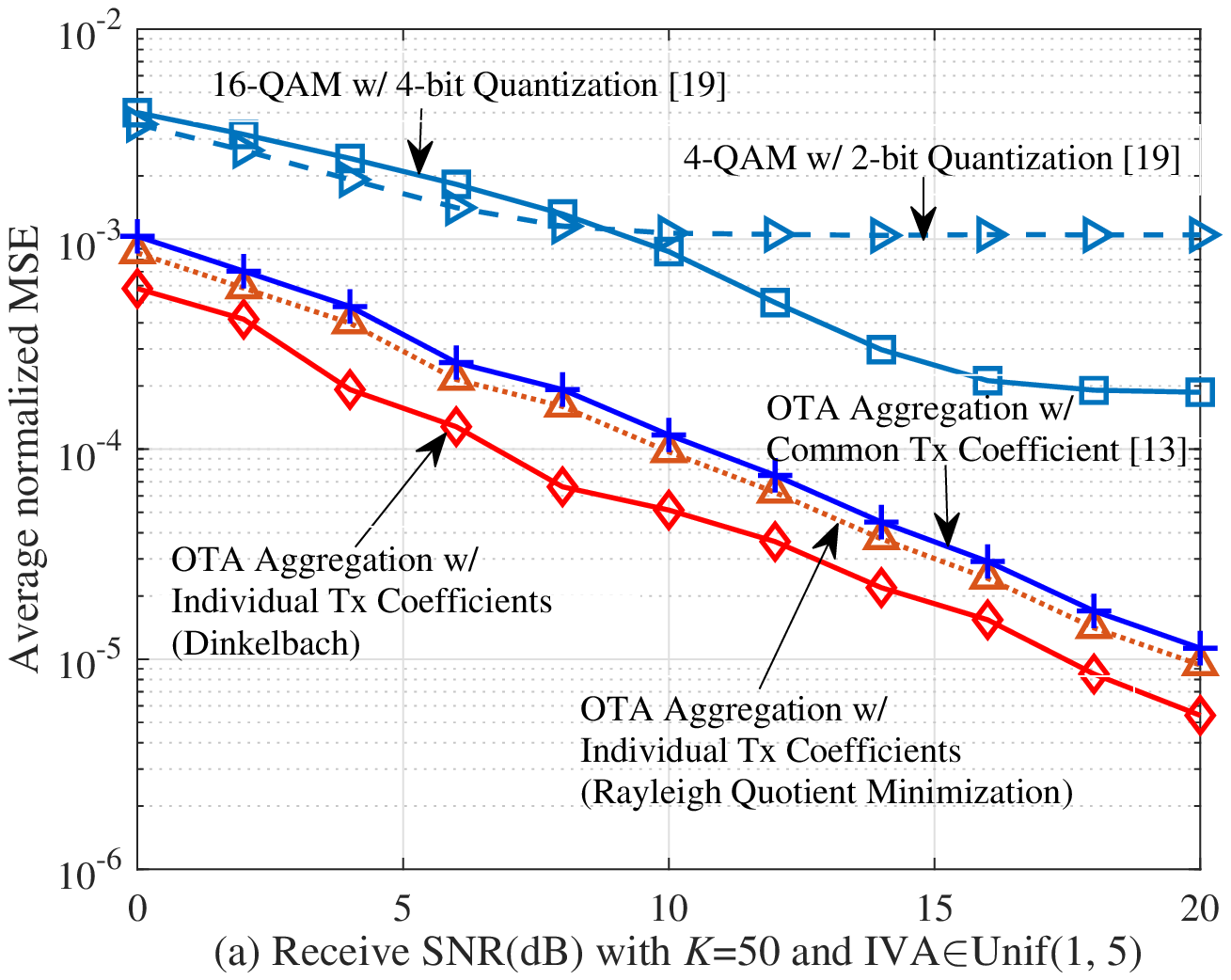}
\end{minipage}
\begin{minipage}[t]{0.48\textwidth}
\centering
\includegraphics[width=3.4in]{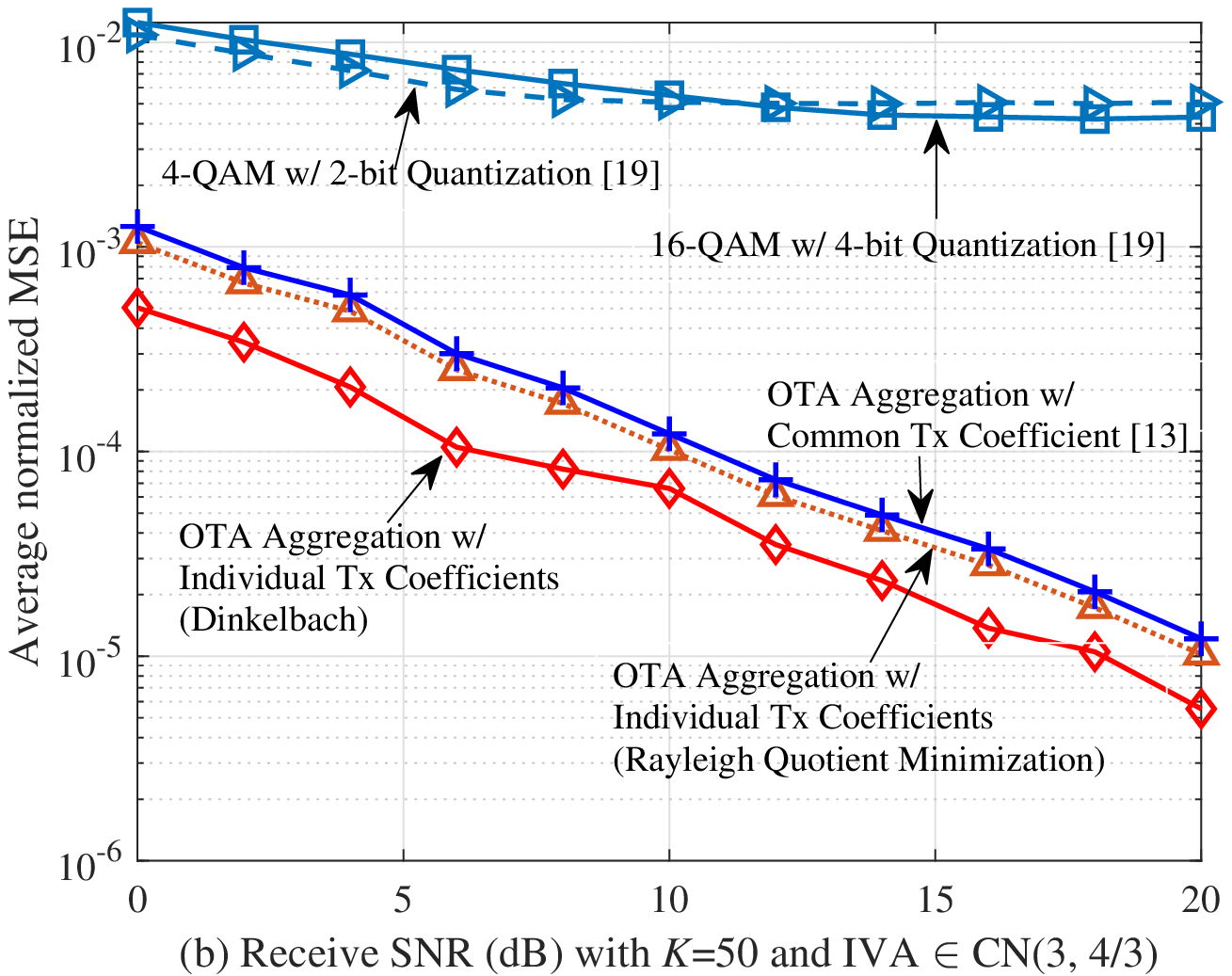}
\end{minipage}
\vspace{-0.4cm}
\caption{The average normalized MSE performance versus the received SNR with the number of WDs $K=50$: (a) the real and imaginary components of the partially aggregated IVA $a^{\rm agg}_i$ are set to be uniformly distributed within the interval $[1,5]$, $\forall i\in{\cal K}$; (b) the partially aggregated IVA is set as $a^{\rm agg}_i\in{\cal CN}(3,4/3)$ for $i\in{\cal K}$.} \label{fig.MSE_vs_RxSNR}
\vspace{-0.5cm}
\end{figure*}

 Fig.~\ref{fig.MSE_vs_RxSNR} shows the average normalized MSE performance versus the received SNR with the number of WDs $K=50$, where the partially aggregated IVAs $a^{\rm agg}_i$ are set to be uniformly distributed within the interval $[1,5]$ and Gaussian distributed ${\cal CN}(3,4/3)$, $\forall i\in{\cal K}$, in Fig.~(\ref{fig.MSE_vs_RxSNR}a) and Fig.~(\ref{fig.MSE_vs_RxSNR}b), respectively. The proposed OTA aggregation scheme based on Dinkelbach's algorithm is observed to achieve a substantial MSE performance gain over the other baseline schemes. This is expected, since the proposed OTA aggregation scheme admits a larger degrees of freedom in optimizing the source WDs' transmit coefficients. Due to the limited degree of freedom in transceiver design, the OTA aggregation scheme with a common transmit coefficient performs inferiorly to the OTA aggregation schemes with independent transmit coefficients at the source WDs. Meanwhile, using the same amount of radio blocks, all three OTA aggregation schemes are observed in Fig.~\ref{fig.MSE_vs_RxSNR} to significantly outperform the digital uncoded QAM schemes, which implies the benefit of analog OTA transmissions in fully utilizing the radio resources. The reason for this is two-fold: first, the digital schemes incur additional quantization error; second, in order for the WDs $\{1,...,K-1\}$ to send their partially aggregated IVAs to WD $K$, the digital schemes require more radio resources than the proposed analog OTA aggregation scheme. When compared with Fig.~\ref{fig.MSE_vs_RxSNR}(a), MSE performance degradation of the digital QAM transmission schemes can be observed in Fig.~\ref{fig.MSE_vs_RxSNR}(b). This is due to the IVA range in the Gaussian distribution being larger than that in the uniform distribution under the same mean and variance, which incurs a larger quantization error. By contrast, the MSE performance of the OTA aggregation schemes only suffer a small variation, which shows a desirable robustness for different IVA distributions.

 \begin{figure*}
\centering
\begin{minipage}[t]{0.48\textwidth}
\centering
\includegraphics[width=3.4in]{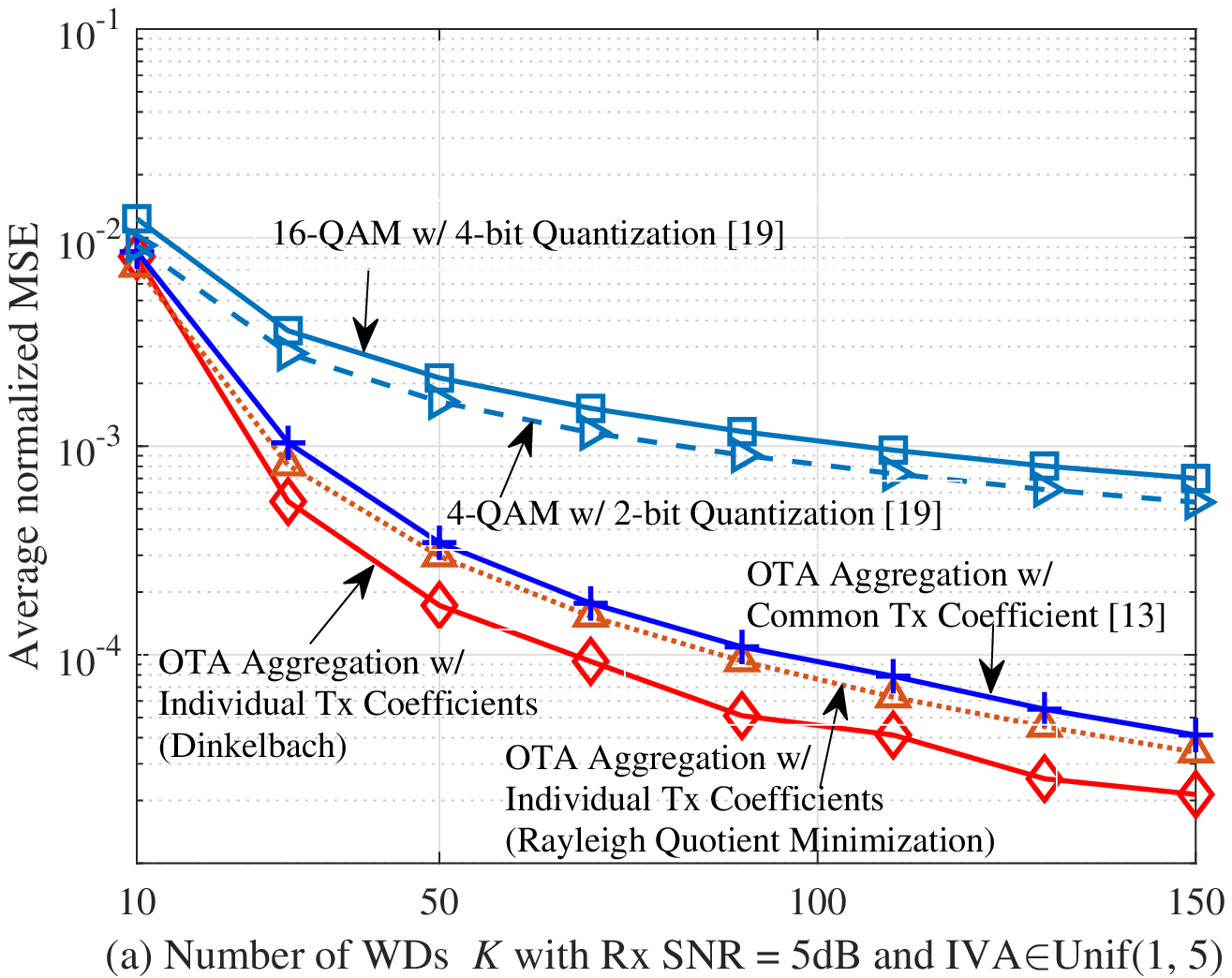}
\end{minipage}
\begin{minipage}[t]{0.48\textwidth}
\centering
\includegraphics[width=3.4in]{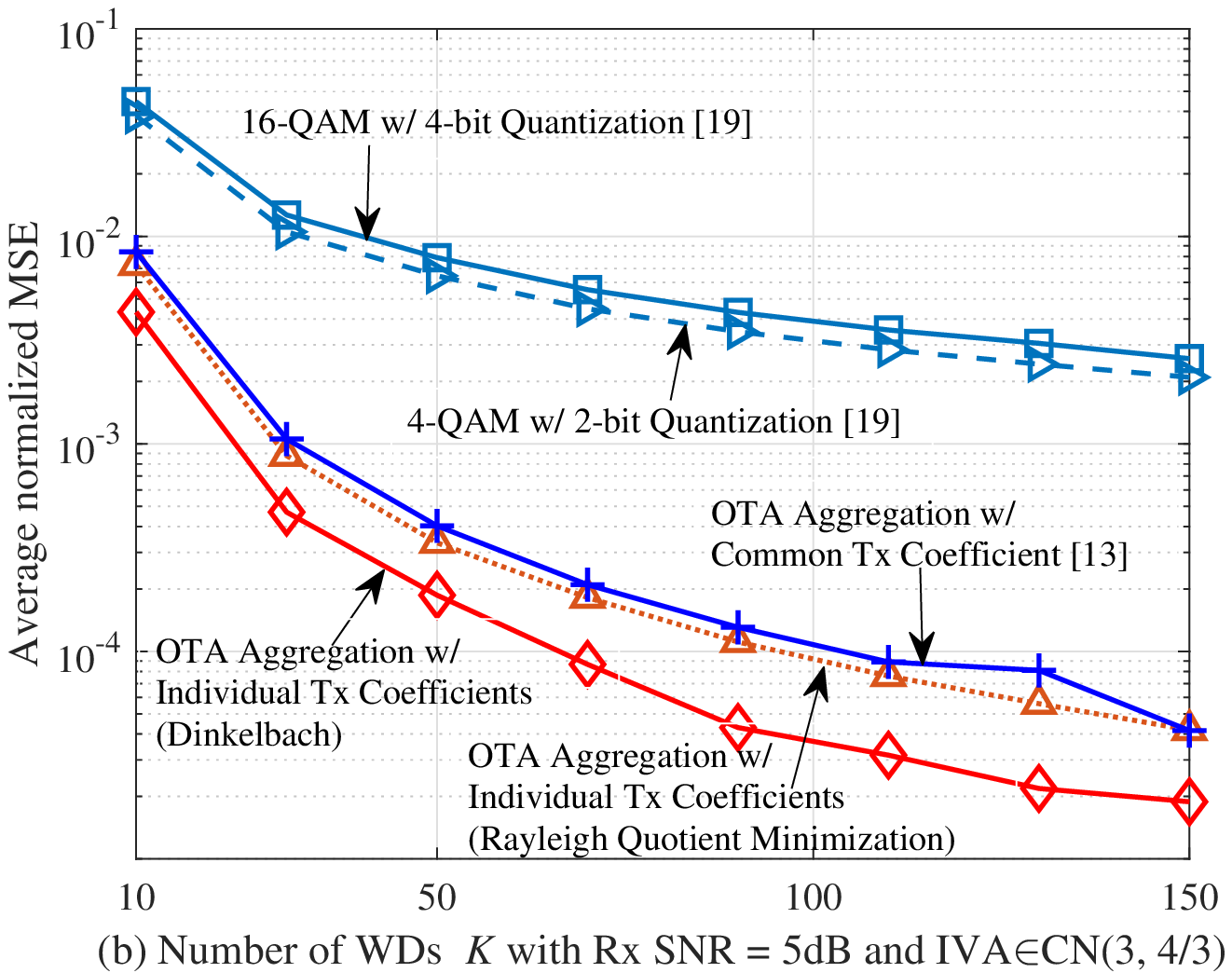}
\end{minipage}
\vspace{-0.4cm}
\caption{The average normalized MSE performance versus the number of WDs $K$ with the received SNR being 5~dB: (a) the real and imaginary components of the partially aggregated IVA $a^{\rm agg}_i$ are set to be uniformly distributed within the interval $[1,5]$, $\forall i\in{\cal K}$; (b) the partially aggregated IVA is set as $a^{\rm agg}_i\in{\cal CN}(3,4/3)$ for $i\in{\cal K}$.} \label{fig.MSE_vs_K}
\vspace{-0.5cm}
\end{figure*}

 Fig.~\ref{fig.MSE_vs_K} shows the average normalized MSE performance versus the number of WDs $K$ with the received SNR being 5~dB, where the partially aggregated IVAs $a^{\rm agg}_i$ are set to be uniformly distributed within the interval $[1,5]$ in Fig.~\ref{fig.MSE_vs_K}(a) and Gaussian distributed ${\cal CN}(3,4/3)$ in Fig.~\ref{fig.MSE_vs_K}(b), $\forall i\in{\cal K}$. It is observed that the proposed multi-level OTA aggregation scheme based on Dinkelbach's algorithm achieves a superior MSE performance over the baseline schemes. As in Fig.~\ref{fig.MSE_vs_RxSNR}, the OTA aggregation scheme with individual transmit coefficients (based on the Rayleigh quotient minimization method) also outperforms that with a common transmit coefficient. Furthermore, the analog OTA aggregation schemes all achieve a substantial MSE performance gain over the digital 4-QAM and 16-QAM schemes, especially as the number of WDs $K$ increases. This is because a large number of WDs $K$ leads to a large amount of radio resources for the digital schemes, while this amount of radio resources for the analog schemes is fixed with an increasing of $K$. Finally, the digital 4-QAM scheme is observed to outperform the 16-QAM scheme in this setup, with a received SNR of 5 dB, which is consistent with the results in Fig.~\ref{fig.MSE_vs_RxSNR}. 
 

\section{Conclusion}\label{section:conclusion}
In this paper, we proposed a new multi-level over-the-air (OTA) aggregation scheme for IVA transmission from multiple source WDs to a destination WD in a wireless D2D MapReduce system. Compared with the conventional digital schemes, the solution has a very low communication resource overhead. By jointly optimizing the source WDs' individual transmit coefficients and the destination WD's receive filter factors, we developed a unified transceiver design that minimizes the computation MSE of the aggregated IVAs subject to the source WDs' individual transmit power constraints. For the case of a common transmit coefficient among the source WDs for OTA aggregation, we derived the closed-form structure of the minimal MSE transceiver design based on the primal decomposition method. Regarding the general case, we transformed the original problem into a quadratic fractional programming problem, and then obtained the (near-) optimal solution by employing Dinkelbach's algorithm Numerical results demonstrated the merits of the proposed design over the existing baseline schemes.


\section*{Appendices}
\subsection{Proof of Lemma~\ref{lemma:opt-gamma}}
 Let $A\triangleq \eta^2|\bm 1^T_{K-1}\bm a|^2+\sigma^2$, $B\triangleq \eta|\bm 1^T_{K-1}\bm a|^2$ and $t={1}/{\gamma(\eta)}$. Based on \eqref{eq.mse-eta-1}, it is yielded that
 \begin{subequations}\label{eq.mse-t}
 \begin{align}
 {\rm MSE}(\eta,\gamma) &= At^2-2Bt+|\bm 1^T_{K-1}\bm a|^2\\
 &= A\Big(t-\frac{B}{A}\Big)^2 - \frac{\eta^2|\bm 1^T_{K-1}\bm a|^4}{\eta^2|\bm 1^T_{K-1}\bm a|^2+\sigma^2} + |\bm 1^T_{K-1}\bm a|^2.
 \end{align}
 \end{subequations}
 
Based on \eqref{eq.mse-t}, the optimal $t^{\rm opt}={1}/{\gamma^{\rm opt}(\eta)}$ to minimize ${\rm MSE}(\eta,\gamma)$ is given as $t^{\rm opt} = B/A$. This then implies that $\frac{1}{\gamma^{\rm opt}(\eta)} = \frac{\eta|\bm 1^T_{K-1}\bm a|^2}{\eta^2|\bm 1_{K-1}^T\bm a|^2+\sigma^2}$. Therefore, under the given feasible $\eta>0$, the optimal $\gamma^{\rm opt}(\eta)$ to minimize ${\rm MSE}(\eta,\gamma)$ is obtained as
$\gamma^{\rm opt}(\eta) =  \frac{\eta^2|\bm 1_{K-1}^T\bm a|^2+\sigma^2}{\eta|\bm 1^T_{K-1}\bm a|^2} = \eta + \frac{\sigma^2}{\eta|\bm 1^T_{K-1}\bm a|^2}$.

\subsection{Proof of Lemma~\ref{lem.mono-mse}}
Define the transmit scaling factor vector $\bm \eta$ as $\bm \eta = \sqrt{\Delta}\tilde{\bm \eta}$, where $\Delta=\|\bm \eta\|^2>0$ and $\tilde{\bm \eta}=\frac{\bm \eta}{\|\bm \eta\|}$. Then, we have ${\rm MSE}(\bm \eta,\gamma^{\rm opt}(\bm \eta)) = |\bm 1_{K-1}^T\bm a|^2 g(\delta)$, where $g(\Delta)$ is defined as
\begin{align}\label{eq.g-delta}
    g(\Delta) = \frac{\Delta{\rm Im}^2(e^{j\theta}\tilde{\bm\eta}^T \bm a )+\sigma^2}{ \Delta|\tilde{\bm \eta}^T \bm a|^2 + \sigma^2},
\end{align}
where $e^{j\theta}=\bm 1_{K-1}^T \bm a/|\bm 1_{K-1}^T \bm a|$. The monotone property of function $g(\Delta)$ can be verified by checking its first-order derivative with respect to $\Delta$. Specifically, with some manipulations, the first-order derivative of $g(\Delta)$ with respect to $\Delta$ is given as
\begin{align}\label{eq.g-prime}
    g^{\prime}(\Delta) &= \frac{ \left({\rm Im}^2(\tilde{\bm\eta}^T \bm a ) -|\tilde{\bm \eta}^T \bm a|^2 \right)\sigma^2}{ (\Delta|\tilde{\bm \eta}^T \bm a|^2 + \sigma^2)^2} = - \frac{ {\rm Re}^2(\tilde{\bm\eta}^T \bm a )\sigma^2}{ (\Delta|\tilde{\bm \eta}^T \bm a|^2 + \sigma^2)^2}\leq 0.
\end{align}
From \eqref{eq.g-prime}, it follows that function $g(\Delta)$ is monotonically decreasing with $\Delta$, and we thus complete the proof of Lemma~\ref{lem.mono-mse}.

\vspace{-0.4cm}
\subsection{Proof of Proposition~\ref{prop:opt-xi}}
By definition, it follows that $0=F(\bm \eta^\star_{\rm ext}, \xi^\star)\leq F(\bm \eta_{\rm ext}, \xi^\star)$. Specifically, we have ${\bm\eta}^{\star T}_{\rm ext}\bm a_3\bm a_3^T {\bm \eta}^\star_{\rm ext} +{\sigma^2} - \xi^{\star}( {\bm\eta}^{\star T}_{\rm ext}(\bm a_1\bm a_1^T + \bm a_2 \bm a_2^T) {\bm \eta}^{\star}_{\rm ext} + {\sigma^2} )=0$ and
\begin{align}
 & {\bm\eta}_{\rm ext}^T\bm a_3\bm a_3^T {\bm \eta}_{\rm ext} +{\sigma^2} - \xi^\star( {\bm\eta}^T_{\rm ext}(\bm a_1\bm a_1^T + \bm a_2 \bm a_2^T) {\bm \eta}_{\rm ext} + {\sigma^2} ) \notag\\
 & \quad\quad\quad \quad \geq {\bm\eta}^{\star T}_{\rm ext}\bm a_3\bm a_3^T {\bm \eta}^\star_{\rm ext} + {\sigma^2} - \xi^{\star}( {\bm\eta}^{\star T}_{\rm ext}(\bm a_1\bm a_1^T + \bm a_2 \bm a_2^T) {\bm \eta}^{\star}_{\rm ext} + {\sigma^2} )   
\end{align}
for any $\bm\eta_{\rm ext}$ satisfying \eqref{eq.tx-p}. Therefore, we have
\begin{align}
& \frac{{\bm\eta}_{\rm ext}^T\bm a_3\bm a_3^T {\bm \eta}_{\rm ext} +{\sigma^2}}{ {\bm\eta}^T_{\rm ext}(\bm a_1\bm a_1^T + \bm a_2 \bm a_2^T) {\bm \eta}_{\rm ext} + {\sigma^2}} \geq \xi^\star = \frac{{\bm\eta}^{\star T}_{\rm ext}\bm a_3\bm a_3^T {\bm \eta}^\star_{\rm ext} +{\sigma^2}}{{\bm\eta}^{\star T}_{\rm ext}(\bm a_1\bm a_1^T + \bm a_2 \bm a_2^T) {\bm \eta}^{\star}_{\rm ext} + {\sigma^2}},
\end{align}
which shows that $\bm \eta_{\rm ext}^\star$ and $\xi^\star$ are the optimal solution and optimal value of $({\cal P}3)$, respectively. 


\end{document}